# Solve the General Constrained Optimal Control Problem with Common Integration Method

Sheng ZHANG and Jin-Mei GAO

(2017.12)

*Abstract:* **Computation of general state- and/or control-constrained Optimal Control Problems (OCPs) is difficult for various constraints, especially the intractable path constraint. For such problems, the theoretical convergence of numerical algorithms is usually not guaranteed, and the right solution may not be successfully obtained. With the recently proposed Variation Evolving Method (VEM), the evolution equations, which guarantee the convergence towards the optimal solution in theory even for the general constrained OCPs, are derived. In particular, the costate-free optimality conditions are established. Besides the analytic expressions of the costates and the Lagrange multipliers adjoining the terminal constraint, the integral equation that determines the Karush-Kuhn-Tucker (KKT) multiplier variable is also derived. Upon the work in this paper, the general constrained OCPs may be transformed to the Initial-value Problems (IVPs) to be solved, with common Ordinary Differential Equation (ODE) numerical integration methods.**

## I. INTRODUCTION

Optimal control theory aims to determine the inputs to a dynamic system that optimize a specified performance index while satisfying constraints on the motion of the system. It is closely related to engineering and has been widely studied [1]. Because of the complexity, Optimal Control Problems (OCPs) are usually solved with numerical methods. Various numerical methods are developed and generally they are divided into two classes, namely, the direct methods and the indirect methods [2]. The direct methods discretize the control or/and state variables to obtain the Nonlinear Programming (NLP) problem, for example, the widely-used direct shooting method [3] and the classic collocation method [4]. These methods are easy to apply, whereas the results obtained are usually suboptimal [5], and the optimal may be infinitely approached. The indirect methods transform the OCP to a Boundary-value Problem (BVP) through the optimality conditions. Typical methods of this type include the well-known indirect shooting method [2] and the novel symplectic method [6]. Although be more precise, the indirect methods often suffer from the significant numerical difficulty due to the ill-conditioning of the Hamiltonian dynamics, that is, the stability of costates dynamics is adverse to that of the states dynamics [7]. The recent development, representatively the Pseudo-spectral (PS) method [8], blends the two types of methods, as it unifies the NLP and the BVP in a dualization view [9]. Such methods inherit the advantages of both types and blur their difference.

Even if there are many numerical methods available, the computation of general state- and/or control-constrained OCPs is still a tough topic today. The complex inequality path constraints result in the daunting optimality conditions in the classic optimal

The authors are with the Computational Aerodynamics Institution, China Aerodynamics Research and Development Center, Mianyang, 621000, China. (e-mail: zszhangshengzs1@outlook.com).

control theory [10]. The indirect multiple shooting method is once the mainstream approach for the optimal solutions of constrained OCPs. However, the utilization of the complex optimality conditions with costates is very user-unfriendly. Moreover, determination of such optimality conditions is still not completely solved [11]. As the development of the direct methods, people prefer to use them because they are much more understandable. Also, the greatly improved efficiency has made the direct methods popular nowadays. However, the theoretic convergence of such algorithms is not well guaranteed for constrained OCPs, even if varieties of promising work has been finished [12][13].

Theories in the control field often enlighten strategies for the optimal control computation, for example, the non-linear variable transformation to reduce the variables [14]. Recently, a Variation Evolving Method (VEM), which is inspired by the continuous-time dynamics stability theory, is proposed for the optimal control computation [15]-[21]. The VEM also synthesizes the direct and indirect methods, but from a new standpoint. The Evolution Partial Differential Equation (EPDE), which describes the evolution of variables towards the optimal solution, is derived from the viewpoint of variation motion, and the optimality conditions will be gradually met under this frame. In Refs. [15] and [16], besides the states and the controls, the costates are also employed in developing the EPDE, and this increases the complexity of the computation. In particular, the time-optimal control problems with control constraint are addressed through that thread. However, it is not widely applicable to the general constrained OCPs. In Ref. [17], a compact version of the VEM that uses only the original variables is proposed. The costate-free optimality conditions are established and the corresponding EPDE is derived for the OCPs with free terminal states. In Refs. [18] and [19], the compact VEM is furthered developed to address the OCPs with terminal Equality Constraints (ECs) and Inequality Constraints (IECs). Normally, under the frame of the compact VEM, the definite conditions for the EPDE are required to be feasible solutions. In Refs. [20] and [21], the Modified EPDE (MEPDE) that is valid even in the infeasible solution domain is proposed to facilitate the computation of the OCPs. In this paper, we further develop the VEM to solve the general constrained OCPs.

Throughout the paper, our work is built upon the assumption that the solution for the optimization problem exists. We do not describe the existing conditions for the purpose of brevity. Relevant researches such as the Filippov-Cesari theorem are documented in Ref. [11]. In the following, first the principle of the VEM and the attributes of IECs in optimization problems are reviewed. Then the VEM for the general constrained OCPs is developed. During this course, the costate-free optimality conditions are established, which uncover the analytic relation of the costates and multipliers in the classic treatment to the state and control variables. Later illustrative examples are solved to verify the effectiveness of the method. Besides, comparison between the derived EPDE and that in Ref. [15] is presented at the end.

## II. Preliminaries

### A. Principle of VEM

The VEM is a newly developed method for the optimal solutions. It is enlightened by the states evolution within the stable continuous-time dynamic system in the control field.

**Lemma 1** [22] (with small adaptation): For a continuous-time autonomous dynamic system like

$$\dot{\boldsymbol{x}} = \boldsymbol{f}(\boldsymbol{x}) \tag{1}$$

where $\boldsymbol{x} \in \mathbb{R}^n$ is the state, $\dot{\boldsymbol{x}} = \dfrac{\mathrm{d}\boldsymbol{x}}{\mathrm{d}t}$ is its time derivative, and $\boldsymbol{f}: \mathbb{R}^n \to \mathbb{R}^n$ is a vector function. Let $\hat{\boldsymbol{x}}$, contained within the domain $\mathbb{D}$, be an equilibrium point that satisfies $\boldsymbol{f}(\hat{\boldsymbol{x}}) = \boldsymbol{0}$. If there exists a continuously differentiable function $V: \mathbb{D} \to \mathbb{R}$ such that

i) $V(\hat{\boldsymbol{x}}) = c$ and $V(\boldsymbol{x}) > c$ in $\mathbb{D} / \{\hat{\boldsymbol{x}}\}$.

ii) $\dot{V}(\boldsymbol{x}) \le 0$ in $\mathbb{D}$ and $\dot{V}(\boldsymbol{x}) < 0$ in $\mathbb{D}/\{\hat{\boldsymbol{x}}\}$.

where $c$ is a constant. Then $\boldsymbol{x} = \hat{\boldsymbol{x}}$ is an asymptotically stable point in $\mathbb{D}$.

Lemma 1 aims to the dynamic system with finite-dimensional states, and it may be directly generalized to the infinite-dimensional case as

**Lemma 2**: For an infinite-dimensional dynamic system described by

$$\frac{\delta \boldsymbol{y}(x)}{\delta t} = \boldsymbol{f}(\boldsymbol{y}, x) \tag{2}$$

or presented equivalently in the Partial Differential Equation (PDE) form as

$$\frac{\partial \boldsymbol{y}(x,t)}{\partial t} = \boldsymbol{f}(\boldsymbol{y}, x) \tag{3}$$

where " $\delta$ " denotes the variation operator and " $\partial$ " denotes the partial differential operator. $t \in \mathbb{R}$ is the time. $x \in \mathbb{R}$ is the independent variable, $\boldsymbol{y}(x) \in \mathbb{R}^n(x)$ is the function vector of $x$, and $\boldsymbol{f}: \mathbb{R}^n(x) \times \mathbb{R} \to \mathbb{R}^n(x)$ is a vector function. Let $\hat{\boldsymbol{y}}(x)$, contained within a certain function set $\mathbb{D}(x)$, is an equilibrium function that satisfies $\boldsymbol{f}(\hat{\boldsymbol{y}}(x), x) = \boldsymbol{0}$. If there exists a continuously differentiable functional $V: \mathbb{D}(x) \to \mathbb{R}$ such that

i) $V\left(\hat{\boldsymbol{y}}(x)\right) = c$ and $V\left(\boldsymbol{y}(x)\right) > c$ in $\mathbb{D}(x)/\{\hat{\boldsymbol{y}}(x)\}$.

ii) $\dot{V}\left(\boldsymbol{y}(x)\right) \le 0$ in $\mathbb{D}(x)$ and $\dot{V}\left(\boldsymbol{y}(x)\right) < 0$ in $\mathbb{D}(x)/\{\hat{\boldsymbol{y}}(x)\}$.

where $c$ is a constant. Then $\boldsymbol{y}(x) = \hat{\boldsymbol{y}}(x)$ is an asymptotically stable solution in $\mathbb{D}(x)$.

In the system dynamics theory, from the stable dynamics, we may construct a monotonously decreasing function (or functional) $V$, which will achieve its minimum when the equilibrium is reached. Inspired by it, now we consider its inverse problem, that is, from a performance index function to derive the dynamics that minimize this performance index, and optimization problems are just the right platform for practice. The optimal solution is analogized to the stable equilibrium of a dynamic system and is anticipated to be obtained in an asymptotically evolving way. Accordingly, a virtual dimension, the variation time $\tau$, is introduced to implement the idea that a variable $\boldsymbol{x}(t)$ evolves to the optimal solution to minimize the performance index within the dynamics governed by the variation dynamic evolution equations (in the form of Eq. (2)). Fig. 1 illustrates the variation evolution process of the VEM in solving the OCP. Through the variation motion, the initial guess of variables will evolve to the optimal solution.

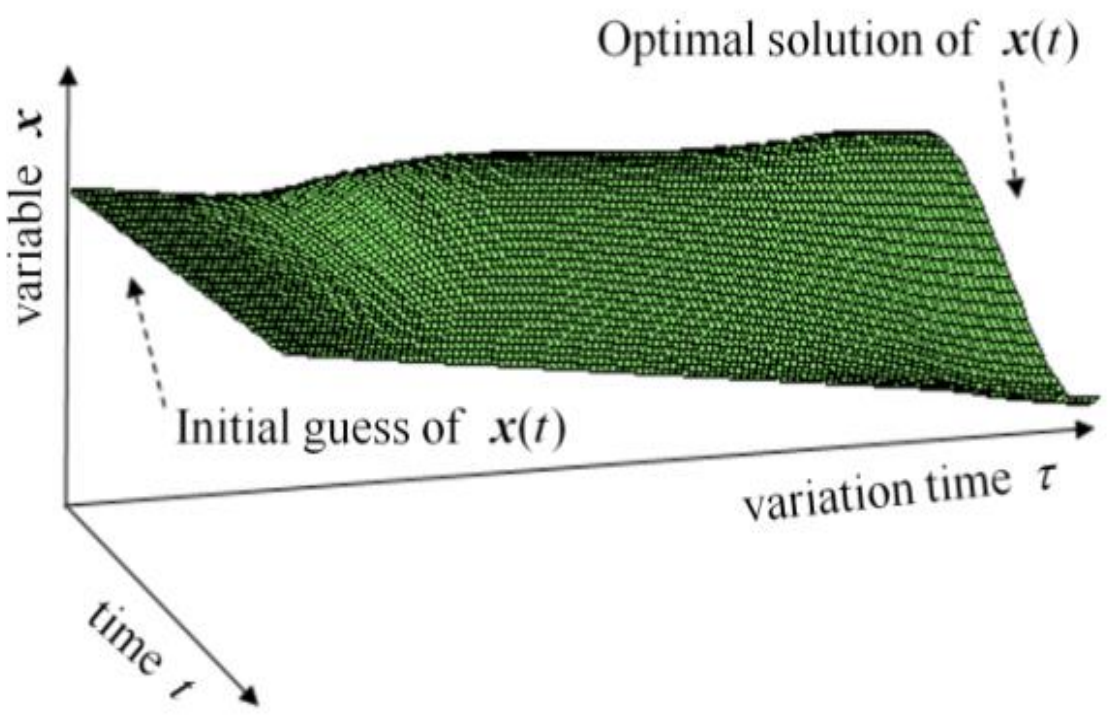

Fig. 1 The illustration of the variable evolving along the variation time $\tau$ in the VEM.

The VEM bred under this idea is demonstrated for the unconstrained calculus-of-variations problems first [15][17]. The variation dynamic evolution equations, derived under the frame of the VEM, may be reformulated as the EPDE and the Evolution Differential Equation (EDE), by replacing the variation operation " $\delta$ " with the partial differential operator " $\partial$ " and the differential operator " d ". Under the dynamics governed by the EPDE, the variables will achieve the optimality conditions gradually. For example, consider the calculus-of-variations problems defined as

$$J = \int_{t_0}^{t_f} F\left(\boldsymbol{y}(t), \dot{\boldsymbol{y}}(t), t\right) \mathrm{d}t \tag{4}$$

where the elements of the variable vector $\boldsymbol{y}(t) \in \mathbb{R}^n(t)$ belong to $C^2[t_0, t_f]$. $t_0$ and $t_f$ are the fixed initial and terminal time, and the boundary conditions are prescribed as $\boldsymbol{y}(t_0) = \boldsymbol{y}_0$ and $\boldsymbol{y}(t_f) = \boldsymbol{y}_f$. The variation dynamic evolution equation obtained with the VEM is

$$\frac{\delta \boldsymbol{y}}{\delta \tau} = -\boldsymbol{K}\left(F_{\boldsymbol{y}} - \frac{\mathrm{d}}{\mathrm{d}t}(F_{\dot{\boldsymbol{y}}})\right) \tag{5}$$

where the column vectors $F_{\boldsymbol{y}} = \frac{\partial F}{\partial \boldsymbol{y}}$ and $F_{\dot{\boldsymbol{y}}} = \frac{\partial F}{\partial \dot{\boldsymbol{y}}}$ are the shorthand notations of partial derivatives, and $\boldsymbol{K}$ is a $n \times n$ dimensional positive-definite gain matrix. Correspondingly, the reformulated EPDE is

$$\frac{\partial \boldsymbol{y}}{\partial \tau} = -\boldsymbol{K}\left(F_{\boldsymbol{y}} - \frac{\partial}{\partial t}(F_{\dot{\boldsymbol{y}}})\right) \tag{6}$$

The equilibrium solution of the EPDE (6) will satisfy the optimality condition, i.e., the Euler-Lagrange equation [23][24]

$$F_{\boldsymbol{y}} - \frac{\mathrm{d}}{\mathrm{d}t}(F_{\dot{\boldsymbol{y}}}) = \boldsymbol{0} \tag{7}$$

Since the right function of the EPDE only depends on the time $t$, it is suitable to be solved via the well-known semi-discrete method in the field of PDE numerical calculation [25]. With the discretization along the normal time dimension, the EPDE is transformed to the finite-dimensional Initial-value Problem (IVP) to be solved, with common Ordinary Differential Equation (ODE) integration methods. Note that the resulting IVP is defined with respect to the variation time $\tau$, not the normal time $t$.

### *B. Active IEC and inactive IEC*

Optimization problems with IECs are more intractable. To search the right evolution equations, Ref. [19] investigated the attributes of ECs and IECs in optimization problems and uncovered the intrinsic relations to the multipliers. Consider the following generalized optimization problem with the performance index as

$$J_g = J_g\left(\boldsymbol{y}(t), \boldsymbol{p}\right) \tag{8}$$

subject to

$$\boldsymbol{g}\left(\boldsymbol{y}(t), \boldsymbol{p}, t\right) = \boldsymbol{0} \qquad t \in \mathbb{T}_E \tag{9}$$

$$\boldsymbol{C}\left(\boldsymbol{y}(t), \boldsymbol{p}, t\right) \le \boldsymbol{0} \qquad t \in \mathbb{T}_I \tag{10}$$

where $\boldsymbol{y}(t) \in \mathbb{R}^{n_y}(t)$ is the optimization variable vector and $\boldsymbol{p} \in \mathbb{R}^{n_p}$ is the optimization parameter vector. Eq. (9) represents the ECs acting in the time set $\mathbb{T}_E$ and Eq. (10) refers to the IECs acting in the time set $\mathbb{T}_I$. Find the optimal solution $\left(\hat{\boldsymbol{y}}(t), \hat{\boldsymbol{p}}\right)$ that minimizes $J_g$, i.e.

$$\left(\hat{\boldsymbol{y}}(t), \hat{\boldsymbol{p}}\right) = \arg\min(J_g) \tag{11}$$

In this general formulation of optimization problems, the ECs (9) are categorized according to their influence to the optimal performance index, denoted by $\hat{J}_g$, and the IECs (10) are classified according to their activeness at the optimal solution, namely

**Definition 1**: Consider a specific time point $t_E \in \mathbb{T}_E$ and reformulate Eq. (9) as

$$\boldsymbol{g}\left(\boldsymbol{y}(t_E), \boldsymbol{p}, t_E\right) = \boldsymbol{a} \tag{12}$$

where $\boldsymbol{a}$ is a right dimensional vector. For the $i$ th component, if there is $\left.\frac{\mathrm{d}\hat{J}_g}{\mathrm{d}a_i}\right|_{a_i=0} \le 0$, then $\left.g_i\right|_{t_E}$ is categorized as a Positive-effect EC; if $\left.\frac{\mathrm{d}\hat{J}_g}{\mathrm{d}a_i}\right|_{a_i=0} > 0$, then $\left.g_i\right|_{t_E}$ is categorized as a Negative-effect EC.

Here in Definition 1, " $|_{a_i=0}$ " means "evaluated at $a_i = 0$" and " $|_{t_E}$ " means "evaluated at $t_E$". In the following, there are similar implications on " $|_{t_0}$ " and " $|_{t_f}$ ".

**Definition 2**: Consider a specific time point $t_I \in \mathbb{T}_I$, for an IEC

$$C_i\left(\boldsymbol{y}(t), \boldsymbol{p}, t_I\right) \le 0 \tag{13}$$

it is said to be an active IEC if

$$C_i\left(\hat{\boldsymbol{y}}(t), \hat{\boldsymbol{p}}, t_I\right) = 0 \tag{14}$$

and said to be an inactive IEC if

$$C_i\left(\hat{\boldsymbol{y}}(t), \hat{\boldsymbol{p}}, t_I\right) < 0 \tag{15}$$

Note that an inactive IEC may be activated for some $\boldsymbol{y}(t)$ and $\boldsymbol{p}$ during the optimization process, but we will not call it an active IEC in the paper. From Definition 2, it is readily to find that strengthening an IEC (13) to be an EC as

$$C_i\left(\boldsymbol{y}(t), \boldsymbol{p}, t_I\right) = 0 \tag{16}$$

the optimal solution will not be changed if this IEC is an active IEC. Also, removing an inactive IEC from the optimization problem, the optimal solution will not be changed either. In order to effectively distinguish the IECs within an optimization problem, we discovered their relations to the strengthened ECs and provided a feasible way. See

**Theorem 1** [19]: The IEC (13) is an active IEC if and only if the strengthened EC (16) is a Positive-effect EC, and the IEC (13) is an inactive IEC if and only if the EC (16) is a Negative-effect EC.

**Theorem 2** [19]: Consider a specific time point $t_E \in \mathbb{T}_E$ and use the Lagrange multiplier $\boldsymbol{\pi}$ to adjoin Eq. (9) with the performance index (8). Then $\left.g_i\right|_{t_E}$ is a Positive-effect EC if and only if $\pi_i \ge 0$. Also, $\left.g_i\right|_{t_E}$ is a Negative-effect EC if and only if $\pi_i < 0$.

Therefore, we may determine the type of an IEC from the multiplier information of its strengthened EC, without the need of substituting optimized solutions into the IEC for verification. In practice, we may first strengthen all IECs to get the corresponding Lagrange multipliers, and then use Theorem 2 to determine their types.

## III. VEM FOR THE GENERAL CONSTRAINED OCPS

### *A. Problem definition*

In this paper, we consider the general constrained OCPs that are defined as

**Problem 1:** Consider performance index of Bolza form

$$J = \varphi(\boldsymbol{x}(t_f), t_f) + \int_{t_0}^{t_f} L\left(\boldsymbol{x}(t), \boldsymbol{u}(t), t\right) \mathrm{d}t \tag{17}$$

subject to the dynamic equation

$$\dot{\boldsymbol{x}} = \boldsymbol{f}(\boldsymbol{x}, \boldsymbol{u}, t) \tag{18}$$

where $t \in \mathbb{R}$ is the time. $\boldsymbol{x} \in \mathbb{R}^n$ are the states and each element is piecewise differentiable. $\boldsymbol{u} \in \mathbb{R}^m$ are the control inputs and each element is piecewise differentiable. The function $L: \mathbb{R}^n \times \mathbb{R}^m \times \mathbb{R} \to \mathbb{R}$ and its first-order partial derivatives are continuous with respect to $\boldsymbol{x}$, $\boldsymbol{u}$ and $t$. The function $\varphi: \mathbb{R}^n \times \mathbb{R} \to \mathbb{R}$ and its first-order partial derivatives are continuous with respect to $\boldsymbol{x}$ and $t$. The vector function $\boldsymbol{f}: \mathbb{R}^n \times \mathbb{R}^m \times \mathbb{R} \to \mathbb{R}^n$ and its first-order partial derivatives are continuous and Lipschitz in $\boldsymbol{x}$, $\boldsymbol{u}$ and $t$. The initial time $t_0$ is fixed and the terminal time $t_f$ is free. The initial and terminal boundary conditions are respectively prescribed as

$$\boldsymbol{x}(t_0) = \boldsymbol{x}_0 \tag{19}$$

$$\boldsymbol{g}\left(\boldsymbol{x}(t_f), t_f\right) = \boldsymbol{0} \tag{20}$$

where $\boldsymbol{g}: \mathbb{R}^n \times \mathbb{R} \to \mathbb{R}^q$ is a $q$ dimensional vector function with continuous first-order partial derivatives. The path constraints are described by

$$\boldsymbol{C}\left(\boldsymbol{x}(t), \boldsymbol{u}(t), t\right) \le \boldsymbol{0} \tag{21}$$

where $\boldsymbol{C}: \mathbb{R}^n \times \mathbb{R}^m \times \mathbb{R} \to \mathbb{R}^r$ is a $r$ dimensional vector function with continuous first-order partial derivatives in its augments. Find the optimal solution $(\hat{\boldsymbol{x}}, \hat{\boldsymbol{u}})$ that minimizes $J$, i.e.

$$(\hat{\boldsymbol{x}}, \hat{\boldsymbol{u}}) = \arg\min(J) \tag{22}$$

The definition of Problem 1 represents a large class of OCPs in engineering. Besides the general Bolza form performance index, for the problems with no terminal constraints or path constraints, they are just the degraded cases of Problem 1; for the situations where the terminal time is fixed, they actually become simpler because now the study requires no determination of the terminal time. Thus, the method developed for Problem 1 may be widely applied and relevant results are of general meaning.

### *B. Derivation of variation dynamic evolution equations*

We first consider the problem within the feasible solution domain $\mathbb{D}_o$, in which any solution satisfies Eqs. (18)-(21). According to the Lyapunov principle, differentiating Eq. (17) with respect to the variation time $\tau$ gives

$$\begin{aligned}
\frac{\delta J}{\delta \tau} &= \varphi_{t_f} \frac{\delta t_f}{\delta \tau} + \varphi_{\boldsymbol{x}_f}{}^{\mathrm{T}} \left( \frac{\delta \boldsymbol{x}(t_f)}{\delta \tau} + (\boldsymbol{f})\Big|_{t_f} \frac{\delta t_f}{\delta \tau} \right) + (L)\Big|_{t_f} \frac{\delta t_f}{\delta \tau} + \int_{t_0}^{t_f} \left( L_{\boldsymbol{x}}{}^{\mathrm{T}} \frac{\delta \boldsymbol{x}}{\delta \tau} + L_{\boldsymbol{u}}{}^{\mathrm{T}} \frac{\delta \boldsymbol{u}}{\delta \tau} \right) \mathrm{d}t \\
&= (\varphi_{t_f} + \varphi_{\boldsymbol{x}_f}{}^{\mathrm{T}} \boldsymbol{f} + L)\Big|_{t_f} \frac{\delta t_f}{\delta \tau} + \varphi_{\boldsymbol{x}_f}{}^{\mathrm{T}} \frac{\delta \boldsymbol{x}(t_f)}{\delta \tau} + \int_{t_0}^{t_f} \left( L_{\boldsymbol{x}}{}^{\mathrm{T}} \frac{\delta \boldsymbol{x}}{\delta \tau} + L_{\boldsymbol{u}}{}^{\mathrm{T}} \frac{\delta \boldsymbol{u}}{\delta \tau} \right) \mathrm{d}t
\end{aligned} \tag{23}$$

where $\varphi_{t_f} := \varphi_t\big|_{t_f}$ and $\varphi_{\boldsymbol{x}_f} := \varphi_{\boldsymbol{x}}\big|_{t_f}$. $\varphi_t = \dfrac{\partial \varphi}{\partial t}$ and $\varphi_{\boldsymbol{x}} = \dfrac{\partial \varphi}{\partial \boldsymbol{x}}$ are the partial derivatives, in the form of scalar and (column) vector, respectively. $L_{\boldsymbol{x}}$ and $L_{\boldsymbol{u}}$ are the partial derivatives. "T" is the transpose operator. For the solutions in $\mathbb{D}_o$, $\dfrac{\delta \boldsymbol{x}}{\delta \tau}$ and $\dfrac{\delta \boldsymbol{u}}{\delta \tau}$ are related because of Eq. (18), and they need to satisfy the following variation equation as

$$\frac{\delta \dot{\boldsymbol{x}}}{\delta \tau} = \boldsymbol{f}_{\boldsymbol{x}} \frac{\delta \boldsymbol{x}}{\delta \tau} + \boldsymbol{f}_{\boldsymbol{u}} \frac{\delta \boldsymbol{u}}{\delta \tau} \tag{24}$$

with the initial condition $\left.\dfrac{\delta \boldsymbol{x}}{\delta \tau}\right|_{t_0} = \boldsymbol{0}$. Note that the Jacobi matrixes $\boldsymbol{f}_{\boldsymbol{x}}$ and $\boldsymbol{f}_{\boldsymbol{u}}$, linearized at the feasible solution $\boldsymbol{x}(t)$ and $\boldsymbol{u}(t)$, are time-dependent. Eq. (24) is a linear time-varying equation and has a zero initial value. Thus according to the linear system theory [26], its solution may be explicitly expressed as

$$\frac{\delta \boldsymbol{x}}{\delta \tau} = \int_{t_0}^{t} \boldsymbol{H}_o(t,s) \frac{\delta \boldsymbol{u}}{\delta \tau}(s)\,\mathrm{d}s \tag{25}$$

where $\boldsymbol{H}_o(t,s)$ is the $n \times m$ dimensional impulse response function corresponding to the specific $\boldsymbol{f}_{\boldsymbol{x}}(t)$ and $\boldsymbol{f}_{\boldsymbol{u}}(t)$, namely

$$\boldsymbol{H}_o(t,s) = \begin{cases} \boldsymbol{\Phi}_o(t,s)\boldsymbol{f}_{\boldsymbol{u}}(s) & t \geq s \\ \boldsymbol{0} & s < t \end{cases} \tag{26}$$

and $\boldsymbol{\Phi}_o(t,s)$ is the $n \times n$ dimensional state transition matrix from time point $s$ to time point $t$, which satisfies

$$\frac{\partial}{\partial t}\boldsymbol{\Phi}_o(t,s) = \boldsymbol{f}_{\boldsymbol{x}}(t)\boldsymbol{\Phi}_o(t,s) \tag{27}$$

In particular,

$$\frac{\delta \boldsymbol{x}(t_f)}{\delta \tau} = \int_{t_0}^{t_f} \boldsymbol{H}_o(t_f,s) \frac{\delta \boldsymbol{u}}{\delta \tau}(s)\,\mathrm{d}s \tag{28}$$

Use Eqs. (25) and (28), and follow the similar derivation as Ref. [17]; we may obtain

$$\frac{\delta J}{\delta \tau} = \left.(\varphi_{t_f} + {\varphi_{\boldsymbol{x}_f}}^{\mathrm{T}} \boldsymbol{f} + L)\right|_{t_f} \frac{\delta t_f}{\delta \tau} + \int_{t_0}^{t_f} {\boldsymbol{p}_{\boldsymbol{u}}}^{\mathrm{T}} \frac{\delta \boldsymbol{u}}{\delta \tau}\,\mathrm{d}t \tag{29}$$

where

$$\boldsymbol{p}_{\boldsymbol{u}}(t) = L_{\boldsymbol{u}} + {\boldsymbol{H}_o}^{\mathrm{T}}(t_f,t)\varphi_{\boldsymbol{x}_f} + \left(\int_t^{t_f} {\boldsymbol{H}_o}^{\mathrm{T}}(\sigma,t) L_{\boldsymbol{x}}(\sigma)\,\mathrm{d}\sigma\right) \tag{30}$$

Now the question of how to find feasible equations for $\dfrac{\delta \boldsymbol{u}}{\delta \tau}$ and $\dfrac{\delta t_f}{\delta \tau}$ arises, which not only guarantee $\dfrac{\delta J}{\delta \tau} \leq 0$ but also satisfy the variation equation of the terminal ECs (20) as

$$\frac{\delta \boldsymbol{g}}{\delta \tau} = \boldsymbol{g}_{\boldsymbol{x}_f} \frac{\delta \boldsymbol{x}(t_f)}{\delta \tau} + \left.\left(\boldsymbol{g}_{\boldsymbol{x}_f} \boldsymbol{f} + \boldsymbol{g}_{t_f}\right)\right|_{t_f} \frac{\delta t_f}{\delta \tau} = \boldsymbol{0} \tag{31}$$

and the variation motion allowed by the path constraints (21) as

$$\frac{\delta C_i}{\delta \tau} = \left(\frac{\partial C_i}{\partial \boldsymbol{x}}\right)^{\mathrm{T}} \frac{\delta \boldsymbol{x}}{\delta \tau} + \left(\frac{\partial C_i}{\partial \boldsymbol{u}}\right)^{\mathrm{T}} \frac{\delta \boldsymbol{u}}{\delta \tau} \leq 0 \qquad t \in \mathbb{T}_i^{P},\ i = 1,2,...,r \tag{32}$$

where $\boldsymbol{g}_{t_f} := \boldsymbol{g}_t\big|_{t_f}$ and $\boldsymbol{g}_{\boldsymbol{x}_f} := \boldsymbol{g}_{\boldsymbol{x}}\big|_{t_f}$. The time set $\mathbb{T}_i^{P}$ is defined for the $i$ th path constraint as

$$\mathbb{T}_i^{P} = \{t \mid C_i(\boldsymbol{x},\boldsymbol{u},t) = 0,\ t \in [t_0, t_f]\} \tag{33}$$

Before answering this question, we introduce the Feasibility-preserving Evolution Optimization Problem (FPEOP) that is defined as

**FPEOP**:

$$\begin{aligned}
&\min \quad J_{t3} = \frac{1}{2}J_{t1} + \frac{1}{2}J_{t2} \\
&s.t. \\
&\frac{\delta \boldsymbol{g}}{\delta \tau} = \boldsymbol{0} \\
&\frac{\delta C_i}{\delta \tau} = 0 \qquad t \in \mathbb{T}_i^{pp},\ i = 1,2,...,r
\end{aligned} \tag{34}$$

where

$$J_{t1} = (\varphi_{t_f} + \varphi_{\boldsymbol{x}_f}{}^{\mathrm{T}} \boldsymbol{f} + L)\Big|_{t_f} \frac{\delta t_f}{\delta \tau} + \int_{t_0}^{t_f} \boldsymbol{p}_{\boldsymbol{u}}{}^{\mathrm{T}} \frac{\delta \boldsymbol{u}}{\delta \tau} \mathrm{d}t \tag{35}$$

$$J_{t2} = \frac{1}{2k_{t_f}} (\frac{\delta t_f}{\delta \tau})^2 + \int_{t_0}^{t_f} \frac{1}{2} (\frac{\delta \boldsymbol{u}}{\delta \tau})^{\mathrm{T}} \boldsymbol{K}^{-1} \frac{\delta \boldsymbol{u}}{\delta \tau} \mathrm{d}t \tag{36}$$

with $\frac{\delta \boldsymbol{u}}{\delta \tau}$ being the optimization variable and $\frac{\delta t_f}{\delta \tau}$ being the optimization parameter. $\boldsymbol{K}$ is a $m \times m$ dimensional positive-definite matrix and $k_{t_f}$ is a positive constant. The time set $\mathbb{T}_i^{pp}$ is a subset of $\mathbb{T}_i^{p}$ defined as

$$\mathbb{T}_i^{pp} = \{t \,|\, C_i(\boldsymbol{x}, \boldsymbol{u}, t) = 0, \frac{\delta C_i}{\delta \tau} \le 0 \text{ is an active IEC, } t \in [t_0, t_f]\} \tag{37}$$

Now the question of determining right $\frac{\delta \boldsymbol{u}}{\delta \tau}$ and $\frac{\delta t_f}{\delta \tau}$ will be answered by the following theorem.

**Theorem 3**: The following variation dynamic evolution equations guarantee that the solution stays in the feasible domain and the change of performance index $\frac{\delta J}{\delta \tau} \le 0$

$$\frac{\delta \boldsymbol{u}}{\delta \tau} = -\boldsymbol{K} \boldsymbol{p}_{\boldsymbol{u}}^{pc} \tag{38}$$

$$\frac{\delta t_f}{\delta \tau} = -k_{t_f} \left( \varphi_{t_f} + \varphi_{\boldsymbol{x}_f}{}^{\mathrm{T}} \boldsymbol{f} + L + \boldsymbol{\pi}^{\mathrm{T}} (\boldsymbol{g}_{\boldsymbol{x}_f} \boldsymbol{f} + \boldsymbol{g}_{t_f}) \right)\Big|_{t_f} \tag{39}$$

where

$$\boldsymbol{p}_{\boldsymbol{u}}^{pc}(t) = \boldsymbol{p}_{\boldsymbol{u}} + \boldsymbol{H}_o{}^{\mathrm{T}}(t_f, t) \boldsymbol{g}_{\boldsymbol{x}_f}{}^{\mathrm{T}} \boldsymbol{\pi} + \boldsymbol{C}_{\boldsymbol{u}}{}^{\mathrm{T}} \boldsymbol{\mu}(t) + \int_t^{t_f} \boldsymbol{H}_o{}^{\mathrm{T}}(s, t) \boldsymbol{C}_{\boldsymbol{x}}{}^{\mathrm{T}}(s) \boldsymbol{\mu}(s) \mathrm{d}s \tag{40}$$

$\boldsymbol{K}$ is the $m \times m$ dimensional positive-definite gain matrix, $k_{t_f}$ is a positive gain constant, and $\boldsymbol{p}_{\boldsymbol{u}}$ is defined in Eq. (30). Assume that the dynamic system satisfies the controllability requirement (See Ref. [27]), then the parameter vector $\boldsymbol{\pi} \in \mathbb{R}^q$ is calculated by

$$\boldsymbol{\pi} = -\boldsymbol{M}^{-1} \boldsymbol{r} \tag{41}$$

The $q \times q$ dimensional matrix $\boldsymbol{M}$ and the $q$ dimensional vector $\boldsymbol{r}$ are

$$\boldsymbol{M} = \boldsymbol{g}_{\boldsymbol{x}_f} \left( \int_{t_0}^{t_f} \boldsymbol{H}_o(t_f, t) \boldsymbol{K} \boldsymbol{H}_o{}^{\mathrm{T}}(t_f, t) \mathrm{d}t \right) \boldsymbol{g}_{\boldsymbol{x}_f}{}^{\mathrm{T}} + k_{t_f} (\boldsymbol{g}_{\boldsymbol{x}_f} \boldsymbol{f} + \boldsymbol{g}_{t_f})(\boldsymbol{g}_{\boldsymbol{x}_f} \boldsymbol{f} + \boldsymbol{g}_{t_f})^{\mathrm{T}} \Big|_{t_f} \tag{42}$$

$$\boldsymbol{r} = \boldsymbol{g}_{\boldsymbol{x}_f} \int_{t_0}^{t_f} \boldsymbol{h}_2(t)\boldsymbol{\mu}(t)\mathrm{d}t + \boldsymbol{h}_1 \tag{43}$$

where

$$\boldsymbol{h}_1 = \boldsymbol{g}_{\boldsymbol{x}_f} \left( \int_{t_0}^{t_f} \boldsymbol{H}_o(t_f,t)\boldsymbol{K}\boldsymbol{p}_{\boldsymbol{u}}\,\mathrm{d}t \right) + k_{t_f}\,(\boldsymbol{g}_{\boldsymbol{x}_f}\boldsymbol{f} + \boldsymbol{g}_{t_f})(\varphi_{t_f} + \varphi_{\boldsymbol{x}_f}{}^{\mathrm{T}}\boldsymbol{f} + L)\Big|_{t_f} \tag{44}$$

$$\boldsymbol{h}_2(t) = \left( \int_{t_0}^{t} \boldsymbol{H}_o(t_f,s)\boldsymbol{K}\boldsymbol{H}_o{}^{\mathrm{T}}(t,s)\,\mathrm{d}s \right)\boldsymbol{C}_{\boldsymbol{x}}{}^{\mathrm{T}} + \boldsymbol{H}_o(t_f,t)\boldsymbol{K}\boldsymbol{C}_{\boldsymbol{u}}{}^{\mathrm{T}} \tag{45}$$

The variable vector $\boldsymbol{\mu}(t) = \begin{bmatrix} \mu_1(t) & \mu_2(t) & \ldots & \mu_r(t) \end{bmatrix}^{\mathrm{T}} \in \mathbb{R}^r(t)$ is determined by

$$\begin{aligned}
&\textit{when} \quad t \notin \mathbb{T}_i^{pp} \qquad (i = 1,2,\ldots,r) \\
&\qquad \mu_i(t) = 0 \\
&\textit{when} \quad t \in \mathbb{T}_i^{pp} \qquad (i = 1,2,\ldots,r) \\
&\qquad \left(\frac{\partial C_i}{\partial \boldsymbol{u}}\right)^{\mathrm{T}} \boldsymbol{K}\boldsymbol{C}_{\boldsymbol{u}}{}^{\mathrm{T}}\boldsymbol{\mu}(t) + \int_{t_0}^{t_f} \boldsymbol{d}_i^W(t,\sigma)\boldsymbol{\mu}(\sigma)\mathrm{d}\sigma + \int_{t_0}^{t} \boldsymbol{d}_i^L(t,\sigma)\boldsymbol{\mu}(\sigma)\,\mathrm{d}\sigma + \int_{t}^{t_f} \boldsymbol{d}_i^R(t,\sigma)\boldsymbol{\mu}(\sigma)\mathrm{d}\sigma + d_i^A(t) = 0
\end{aligned} \tag{46}$$

with

$$\boldsymbol{d}_i^W(t,\sigma) = -\left(\frac{\partial C_i}{\partial \boldsymbol{x}}(t)\right)^{\mathrm{T}} \left( \int_{t_0}^{t} \boldsymbol{H}_o(t,s)\boldsymbol{K}\boldsymbol{H}_o{}^{\mathrm{T}}(t_f,s)\,\mathrm{d}s \right) \boldsymbol{g}_{\boldsymbol{x}_f}{}^{\mathrm{T}}\boldsymbol{M}^{-1}\boldsymbol{g}_{\boldsymbol{x}_f}\boldsymbol{h}_2(\sigma) - \left(\frac{\partial C_i}{\partial \boldsymbol{u}}(t)\right)^{\mathrm{T}} \boldsymbol{K}\boldsymbol{H}_o{}^{\mathrm{T}}(t_f,t)\boldsymbol{g}_{\boldsymbol{x}_f}{}^{\mathrm{T}}\boldsymbol{M}^{-1}\boldsymbol{g}_{\boldsymbol{x}_f}\boldsymbol{h}_2(\sigma) \tag{47}$$

$$\boldsymbol{d}_i^L(t,\sigma) = \left(\frac{\partial C_i}{\partial \boldsymbol{x}}(t)\right)^{\mathrm{T}} \left( \int_{t_0}^{\sigma} \boldsymbol{H}_o(t,s)\boldsymbol{K}\boldsymbol{H}_o{}^{\mathrm{T}}(\sigma,s)\,\mathrm{d}s \right) \boldsymbol{C}_{\boldsymbol{x}}{}^{\mathrm{T}}(\sigma) + \left(\frac{\partial C_i}{\partial \boldsymbol{x}}(t)\right)^{\mathrm{T}} \boldsymbol{H}_o(t,\sigma)\boldsymbol{K}\boldsymbol{C}_{\boldsymbol{u}}{}^{\mathrm{T}}(\sigma) \tag{48}$$

$$\boldsymbol{d}_i^R(t,\sigma) = \left(\frac{\partial C_i}{\partial \boldsymbol{x}}(t)\right)^{\mathrm{T}} \left( \int_{t_0}^{t} \boldsymbol{H}_o(t,s)\boldsymbol{K}\boldsymbol{H}_o{}^{\mathrm{T}}(\sigma,s)\,\mathrm{d}s \right) \boldsymbol{C}_{\boldsymbol{x}}{}^{\mathrm{T}}(\sigma) + \left(\frac{\partial C_i}{\partial \boldsymbol{u}}(t)\right)^{\mathrm{T}} \boldsymbol{K}\boldsymbol{H}_o{}^{\mathrm{T}}(\sigma,t)\boldsymbol{C}_{\boldsymbol{x}}{}^{\mathrm{T}}(\sigma) \tag{49}$$

$$d_i^A(t) = \left(\frac{\partial C_i}{\partial \boldsymbol{x}}\right)^{\mathrm{T}} \int_{t_0}^{t} \boldsymbol{H}_o(t,s)\boldsymbol{K}\boldsymbol{p}_{\boldsymbol{u}}\,\mathrm{d}s - \left(\frac{\partial C_i}{\partial \boldsymbol{x}}\right)^{\mathrm{T}} \left( \int_{t_0}^{t} \boldsymbol{H}_o(t,s)\boldsymbol{K}\boldsymbol{H}_o{}^{\mathrm{T}}(t_f,s)\,\mathrm{d}s \right) \boldsymbol{g}_{\boldsymbol{x}_f}{}^{\mathrm{T}}\boldsymbol{M}^{-1}\boldsymbol{h}_1 + \left(\frac{\partial C_i}{\partial \boldsymbol{u}}\right)^{\mathrm{T}} \boldsymbol{K}\boldsymbol{p}_{\boldsymbol{u}} - \left(\frac{\partial C_i}{\partial \boldsymbol{u}}\right)^{\mathrm{T}} \boldsymbol{K}\boldsymbol{H}_o{}^{\mathrm{T}}(t_f,t)\boldsymbol{g}_{\boldsymbol{x}_f}{}^{\mathrm{T}}\boldsymbol{M}^{-1}\boldsymbol{h}_1 \tag{50}$$

Moreover, under the evolution equations (38) and (39), $\frac{\delta J}{\delta \tau} = 0$ occurs only when

$$\boldsymbol{p}_{\boldsymbol{u}}^{pc}(t) = \boldsymbol{0} \tag{51}$$

$$\left( \varphi_{t_f} + \varphi_{\boldsymbol{x}_f}{}^{\mathrm{T}}\boldsymbol{f} + L + \boldsymbol{\pi}^{\mathrm{T}}(\boldsymbol{g}_{\boldsymbol{x}_f}\boldsymbol{f} + \boldsymbol{g}_{t_f}) \right)\Big|_{t_f} = 0 \tag{52}$$

**Proof**: We will derive Eqs. (38) and (39) though the optimization theory. Reformulate Eq. (29) as a constrained optimization problem (use $J_{t1}$ to denote the performance index as defined in Eq. (35)) subject to constraints (31) and (32). Note that now $\frac{\delta \boldsymbol{u}}{\delta \tau}$ is the optimization variable and $\frac{\delta t_f}{\delta \tau}$ is the optimization parameter. However, since the minimum of this optimization problem may be negative infinity, to penalize too large optimization variable (parameter), we introduce another performance index $J_{t2}$ as defined in Eq. (36) to formulate a Multi-objective Optimization Problem (MOP) as

$$\begin{aligned}
&\min(J_{t1}, J_{t2}) \\
&s.t. \\
&\frac{\delta \boldsymbol{g}}{\delta \tau} = \boldsymbol{0} \\
&\frac{\delta C_i}{\delta \tau} \le 0 \qquad t \in \mathbb{T}_i^{p},\ i = 1,2,\ldots,r
\end{aligned} \tag{53}$$

We use the weighting method to solve the Pareto optimal solution of this MOP, and the resulting performance index is

$$J_{t3}=aJ_{t1}+bJ_{t2} \tag{54}$$

where $a\geq 0$, $b\geq 0$ and $a+b=1$. When $a=1,\ b=0$, we get a solution that minimizes $J_{t1}$. When $a=0,\ b=1$, we get a solution that minimizes $J_{t2}$. Otherwise, we get a compromising solution. For this MOP, obviously in the case of $a=0,\ b=1$, the Pareto optimal solution is that $\frac{\delta t_f}{\delta\tau}=0$ and $\frac{\delta\boldsymbol{u}}{\delta\tau}=\boldsymbol{0}$, and now the value of performance indexes are $J_{t1}=0$ and $J_{t2}=0$. For any other cases, the compromising solution guarantees that $J_{t1}\leq 0$. Set $a=\frac{1}{2}$, $b=\frac{1}{2}$, and because the inactive IECs may be removed and the active IECs may be strengthened without changing the optimal solution, then we have the FPEOP defined in Eq. (34).

Introduce the Lagrange multiplier parameter $\boldsymbol{\pi}\in\mathbb{R}^q$ and the KKT multiplier variable $\boldsymbol{\mu}\in\mathbb{R}^r$ to adjoin the constraints, we may get the unconstrained optimization problem from the FPEOP as

$$\begin{aligned}J_{t4}&=\frac{1}{2}J_{t1}+\frac{1}{2}J_{t2}+\frac{1}{2}\boldsymbol{\pi}^{\mathrm{T}}\frac{\delta\boldsymbol{g}}{\delta\tau}+\frac{1}{2}\int_{t_0}^{t_f}\boldsymbol{\mu}^{\mathrm{T}}\frac{\delta\boldsymbol{C}}{\delta\tau}\mathrm{d}t\\&=\frac{1}{2}J_{t1}+\frac{1}{2}J_{t2}+\frac{1}{2}\boldsymbol{\pi}^{\mathrm{T}}\left(\boldsymbol{g}_{\boldsymbol{x}_f}\int_{t_0}^{t_f}\boldsymbol{H}_o(t_f,t)\frac{\delta\boldsymbol{u}}{\delta\tau}(t)\mathrm{d}t+\boldsymbol{g}_{\boldsymbol{x}_f}(\boldsymbol{f})\big|_{t_f}\frac{\delta t_f}{\delta\tau}+\boldsymbol{g}_{t_f}\frac{\delta t_f}{\delta\tau}\right)+\frac{1}{2}\int_{t_0}^{t_f}\boldsymbol{\mu}^{\mathrm{T}}\left(\boldsymbol{C}_{\boldsymbol{x}}\int_{t_0}^{t}\boldsymbol{H}_o(t,s)\frac{\delta\boldsymbol{u}}{\delta\tau}(s)\mathrm{d}s+\boldsymbol{C}_{\boldsymbol{u}}\frac{\delta\boldsymbol{u}}{\delta\tau}\right)\mathrm{d}t\end{aligned} \tag{55}$$

with

$$\mu_i(t)=0\quad when\quad t\notin\mathbb{T}_i^{pp}\qquad(i=1,2,...,r) \tag{56}$$

By exchanging the order in the double integral and the symbols $t\rightleftarrows s$, there is

$$\int_{t_0}^{t_f}\boldsymbol{\mu}^{\mathrm{T}}\boldsymbol{C}_{\boldsymbol{x}}\left(\int_{t_0}^{t}\boldsymbol{H}_o(t,s)\frac{\delta\boldsymbol{u}}{\delta\tau}(s)\mathrm{d}s\right)\mathrm{d}t=\int_{t_0}^{t_f}\left(\int_{t}^{t_f}\boldsymbol{\mu}^{\mathrm{T}}\boldsymbol{C}_{\boldsymbol{x}}\boldsymbol{H}_o(s,t)\mathrm{d}s\right)\frac{\delta\boldsymbol{u}}{\delta\tau}(t)\mathrm{d}t \tag{57}$$

Thus

$$J_{t4}=\frac{1}{2}J_{t1}+\frac{1}{2}J_{t2}+\frac{1}{2}\boldsymbol{\pi}^{\mathrm{T}}\left(\boldsymbol{g}_{\boldsymbol{x}_f}\int_{t_0}^{t_f}\boldsymbol{H}_o(t_f,t)\frac{\delta\boldsymbol{u}}{\delta\tau}(t)\mathrm{d}t+\boldsymbol{g}_{\boldsymbol{x}_f}(\boldsymbol{f})\big|_{t_f}\frac{\delta t_f}{\delta\tau}+\boldsymbol{g}_{t_f}\frac{\delta t_f}{\delta\tau}\right)+\frac{1}{2}\int_{t_0}^{t_f}\left(\boldsymbol{\mu}^{\mathrm{T}}\boldsymbol{C}_{\boldsymbol{u}}+\int_{t}^{t_f}\boldsymbol{\mu}^{\mathrm{T}}\boldsymbol{C}_{\boldsymbol{x}}\boldsymbol{H}_o(s,t)\mathrm{d}s\right)\frac{\delta\boldsymbol{u}}{\delta\tau}(t)\mathrm{d}t \tag{58}$$

Use the fist-order optimality conditions, i.e., $\frac{\partial J_{t4}}{\partial\left(\frac{\delta\boldsymbol{u}}{\delta\tau}\right)}=\boldsymbol{0}$ and $\frac{\partial J_{t4}}{\partial\left(\frac{\delta t_f}{\delta\tau}\right)}=0$, we may get Eqs. (38) and (39).

Substitute Eqs. (38) and (39) into Eq. (31); we have

$$\begin{aligned}&\boldsymbol{g}_{\boldsymbol{x}_f}\int_{t_0}^{t_f}\boldsymbol{H}_o(t_f,t)\boldsymbol{K}\left(\boldsymbol{p}_{\boldsymbol{u}}+\boldsymbol{H}_o^{\mathrm{T}}(t_f,t)\boldsymbol{g}_{\boldsymbol{x}_f}^{\mathrm{T}}\boldsymbol{\pi}+\int_{t}^{t_f}\boldsymbol{H}_o^{\mathrm{T}}(s,t)\boldsymbol{C}_{\boldsymbol{x}}^{\mathrm{T}}(s)\boldsymbol{\mu}(s)\mathrm{d}s+\boldsymbol{C}_{\boldsymbol{u}}^{\mathrm{T}}\boldsymbol{\mu}(t)\right)\mathrm{d}t\\&+k_{t_f}\left(\boldsymbol{g}_{\boldsymbol{x}_f}\boldsymbol{f}+\boldsymbol{g}_{t_f}\right)\left(\varphi_{t_f}+\varphi_{\boldsymbol{x}_f}^{\mathrm{T}}\boldsymbol{f}+L+\boldsymbol{\pi}^{\mathrm{T}}(\boldsymbol{g}_{\boldsymbol{x}_f}\boldsymbol{f}+\boldsymbol{g}_{t_f})\right)\Big|_{t_f}=\boldsymbol{0}\end{aligned} \tag{59}$$

Again with the technique of exchanging the order in the double integral and the symbols $t\rightleftarrows s$, there is

$$\int_{t_0}^{t_f}\boldsymbol{H}_o(t_f,t)\boldsymbol{K}\left(\int_{t}^{t_f}\boldsymbol{H}_o^{\mathrm{T}}(s,t)\boldsymbol{C}_{\boldsymbol{x}}^{\mathrm{T}}(s)\boldsymbol{\mu}(s)\mathrm{d}s\right)\mathrm{d}t=\int_{t_0}^{t_f}\left(\int_{t_0}^{t}\boldsymbol{H}_o(t_f,s)\boldsymbol{K}\boldsymbol{H}_o^{\mathrm{T}}(t,s)\mathrm{d}s\right)\boldsymbol{C}_{\boldsymbol{x}}^{\mathrm{T}}(t)\boldsymbol{\mu}(t)\mathrm{d}t \tag{60}$$

With further deduction, we have

$$\begin{aligned}&\left(\boldsymbol{g}_{\boldsymbol{x}_f}\left(\int_{t_0}^{t_f}\boldsymbol{H}_o(t_f,t)\boldsymbol{K}\boldsymbol{H}_o^{\mathrm{T}}(t_f,t)\mathrm{d}t\right)\boldsymbol{g}_{\boldsymbol{x}_f}^{\mathrm{T}}+k_{t_f}(\boldsymbol{g}_{\boldsymbol{x}_f}\boldsymbol{f}+\boldsymbol{g}_{t_f})(\boldsymbol{g}_{\boldsymbol{x}_f}\boldsymbol{f}+\boldsymbol{g}_{t_f})^{\mathrm{T}}\Big|_{t_f}\right)\boldsymbol{\pi}\\&=-\boldsymbol{g}_{\boldsymbol{x}_f}\int_{t_0}^{t_f}\boldsymbol{h}_2(t)\boldsymbol{\mu}(t)\mathrm{d}t-\boldsymbol{h}_1\end{aligned} \tag{61}$$

where $\boldsymbol{h}_1$ and $\boldsymbol{h}_2(t)$ are defined in Eqs. (44) and (45), respectively. Thus, with the definition of $\boldsymbol{M}$ in Eq. (42) and $\boldsymbol{r}$ in Eq. (43), Eq. (61) that determines $\boldsymbol{\pi}$ is simplified as

$$\boldsymbol{M}\boldsymbol{\pi} = -\boldsymbol{r} \tag{62}$$

Regarding this linear equation, assuming that the control satisfies the controllability requirement [27], then the solution is guaranteed. Thus the parameter $\boldsymbol{\pi}$ may be calculated by Eq. (41).

For $t \in \mathbb{T}_i^{pp}$ $(i = 1,2,...,r)$ , there is $\frac{\delta C_i}{\delta \tau} = 0$ , i.e.

$$(\frac{\partial C_i}{\partial \boldsymbol{x}})^{\mathrm{T}} \int_{t_0}^{t} \boldsymbol{H}_o(t,s) \frac{\delta \boldsymbol{u}}{\delta \tau}(s)\,\mathrm{d}s + (\frac{\partial C_i}{\partial \boldsymbol{u}})^{\mathrm{T}} \frac{\delta \boldsymbol{u}}{\delta \tau} = 0 \tag{63}$$

Substituting Eq. (38) into Eq. (63) gives

$$\begin{aligned}
&(\frac{\partial C_i}{\partial \boldsymbol{x}})^{\mathrm{T}} \int_{t_0}^{t} \boldsymbol{H}_o(t,s)\boldsymbol{K}\left( \boldsymbol{p}_{\boldsymbol{u}} + \boldsymbol{H}_o^{\mathrm{T}}(t_f,s)\boldsymbol{g}_{\boldsymbol{x}_f}{}^{\mathrm{T}}\boldsymbol{\pi} + \int_{s}^{t_f} \boldsymbol{H}_o^{\mathrm{T}}(\sigma,s)\boldsymbol{C}_{\boldsymbol{x}}{}^{\mathrm{T}}(\sigma)\boldsymbol{\mu}(\sigma)\mathrm{d}\sigma + \boldsymbol{C}_{\boldsymbol{u}}{}^{\mathrm{T}}\boldsymbol{\mu}(s) \right)\mathrm{d}s \\
&+ (\frac{\partial C_i}{\partial \boldsymbol{u}})^{\mathrm{T}} \boldsymbol{K}\left( \boldsymbol{p}_{\boldsymbol{u}} + \boldsymbol{H}_o^{\mathrm{T}}(t_f,t)\boldsymbol{g}_{\boldsymbol{x}_f}{}^{\mathrm{T}}\boldsymbol{\pi} + \int_{t}^{t_f} \boldsymbol{H}_o(s,t)^{\mathrm{T}}\boldsymbol{C}_{\boldsymbol{x}}{}^{\mathrm{T}}(s)\boldsymbol{\mu}(s)\mathrm{d}s + \boldsymbol{C}_{\boldsymbol{u}}{}^{\mathrm{T}}\boldsymbol{\mu}(t) \right) = 0
\end{aligned} \tag{64}$$

Substitute Eq. (41) in and use

$$\begin{aligned}
&\int_{t_0}^{t} \boldsymbol{H}_o(t,s)\boldsymbol{K}\left( \int_{s}^{t_f} \boldsymbol{H}_o^{\mathrm{T}}(\sigma,s)\boldsymbol{C}_{\boldsymbol{x}}{}^{\mathrm{T}}(\sigma)\boldsymbol{\mu}(\sigma)\mathrm{d}\sigma \right)\mathrm{d}s \\
&= \int_{t_0}^{t} \left( \int_{t_0}^{\sigma} \boldsymbol{H}_o(t,s)\boldsymbol{K}\boldsymbol{H}_o^{\mathrm{T}}(\sigma,s)\,\mathrm{d}s \right)\boldsymbol{C}_{\boldsymbol{x}}{}^{\mathrm{T}}(\sigma)\boldsymbol{\mu}(\sigma)\mathrm{d}\sigma + \int_{t}^{t_f} \left( \int_{t_0}^{t} \boldsymbol{H}_o(t,s)\boldsymbol{K}\boldsymbol{H}_o^{\mathrm{T}}(\sigma,s)\,\mathrm{d}s \right)\boldsymbol{C}_{\boldsymbol{x}}{}^{\mathrm{T}}(\sigma)\boldsymbol{\mu}(\sigma)\mathrm{d}\sigma
\end{aligned} \tag{65}$$

$$\int_{t_0}^{t} \boldsymbol{H}_o(t,s)\boldsymbol{K}\boldsymbol{H}_o^{\mathrm{T}}(t_f,s)\boldsymbol{g}_{\boldsymbol{x}_f}{}^{\mathrm{T}}\boldsymbol{M}^{-1}\left( \boldsymbol{g}_{\boldsymbol{x}_f} \int_{t_0}^{t_f} \boldsymbol{h}_2(\sigma)\boldsymbol{\mu}(\sigma)\mathrm{d}\sigma \right)\mathrm{d}s = \int_{t_0}^{t_f} \left( \int_{t_0}^{t} \boldsymbol{H}_o(t,s)\boldsymbol{K}\boldsymbol{H}_o^{\mathrm{T}}(t_f,s)\,\mathrm{d}s \right)\boldsymbol{g}_{\boldsymbol{x}_f}{}^{\mathrm{T}}\boldsymbol{M}^{-1}\boldsymbol{g}_{\boldsymbol{x}_f}\boldsymbol{h}_2(\sigma)\boldsymbol{\mu}(\sigma)\mathrm{d}\sigma \tag{66}$$

Then we have

$$(\frac{\partial C_i}{\partial \boldsymbol{u}})^{\mathrm{T}} \boldsymbol{K}\boldsymbol{C}_{\boldsymbol{u}}{}^{\mathrm{T}}\boldsymbol{\mu}(t) + \int_{t_0}^{t_f} \boldsymbol{d}_i^{W}(t,\sigma)\boldsymbol{\mu}(\sigma)\mathrm{d}\sigma + \int_{t_0}^{t} \boldsymbol{d}_i^{L}(t,\sigma)\boldsymbol{\mu}(\sigma)\,\mathrm{d}\sigma + \int_{t}^{t_f} \boldsymbol{d}_i^{R}(t,\sigma)\boldsymbol{\mu}(\sigma)\mathrm{d}\sigma + d_i^{A}(t) = 0 \tag{67}$$

where the $1 \times m$ dimensional matrixes $\boldsymbol{d}_i^{W}(t,\sigma)$ , $\boldsymbol{d}_i^{L}(t,\sigma)$ , $\boldsymbol{d}_i^{R}(t,\sigma)$ and the scalar quantity $d_i^{A}(t)$ are defined in Eqs. (47)-(50).

Furthermore, Eq. (29) may be reformulated as

$$\begin{aligned}
\frac{\delta J}{\delta \tau} =& \left( \varphi_{t_f} + \varphi_{\boldsymbol{x}_f}{}^{\mathrm{T}}\boldsymbol{f} + L + \boldsymbol{\pi}^{\mathrm{T}}(\boldsymbol{g}_{\boldsymbol{x}_f}\boldsymbol{f} + \boldsymbol{g}_{t_f}) \right)\Big|_{t_f} \frac{\delta t_f}{\delta \tau} + \int_{t_0}^{t_f} \left( \boldsymbol{p}_{\boldsymbol{u}}^{pc} \right)^{\mathrm{T}} \frac{\delta \boldsymbol{u}}{\delta \tau}\mathrm{d}t \\
&- \int_{t_0}^{t_f} \boldsymbol{\mu}^{\mathrm{T}}\left( \boldsymbol{C}_{\boldsymbol{x}} \int_{t_0}^{t} \boldsymbol{H}_o(t,s)\frac{\delta \boldsymbol{u}}{\delta \tau}(s)\,\mathrm{d}s + \boldsymbol{C}_{\boldsymbol{u}}\frac{\delta \boldsymbol{u}}{\delta \tau} \right)\mathrm{d}t \\
&- \boldsymbol{\pi}^{\mathrm{T}}\left( \boldsymbol{g}_{\boldsymbol{x}_f} \int_{t_0}^{t_f} \boldsymbol{H}_o(t_f,t)\frac{\delta \boldsymbol{u}}{\delta \tau}\mathrm{d}t + (\boldsymbol{g}_{\boldsymbol{x}_f}\boldsymbol{f} + \boldsymbol{g}_{t_f})\Big|_{t_f} \frac{\delta t_f}{\delta \tau} \right)
\end{aligned} \tag{68}$$

Since under Eqs. (38) and (39), the last two terms in the right part of Eq. (68) vanish. Then

$$\frac{\delta J}{\delta \tau} = -k_{t_f}\left( \varphi_{t_f} + \varphi_{\boldsymbol{x}_f}{}^{\mathrm{T}}\boldsymbol{f} + L + \boldsymbol{\pi}^{\mathrm{T}}(\boldsymbol{g}_{\boldsymbol{x}_f}\boldsymbol{f} + \boldsymbol{g}_{t_f}) \right)^2\Big|_{t_f} - \int_{t_0}^{t_f} (\boldsymbol{p}_{\boldsymbol{u}}^{pc})^{\mathrm{T}}\boldsymbol{K}\boldsymbol{p}_{\boldsymbol{u}}^{pc}\,\mathrm{d}t \tag{69}$$

This means $\frac{\delta J}{\delta \tau} \leq 0$ and $\frac{\delta J}{\delta \tau} = 0$ occurs only when Eqs. (51) and (52) hold. ■

**Remark 1**: For the optimal solution, there is $\mathbb{T}_i^{pp} = \mathbb{T}_i^{p}$ $(i = 1,2,...,r)$ . The optimal value of $\boldsymbol{\mu}(t)$ (corresponding to the right $\mathbb{T}_i^{pp}$ ) and the optimal value of $\boldsymbol{\pi}$ satisfy the gain-independent equations as

$$
\begin{aligned}
&\textit{when} \quad t \notin \mathbb{T}_i^{pp} \qquad (i = 1, 2, \ldots, r) \\
&\qquad \mu_i(t) = 0 \\
&\textit{when} \quad t \in \mathbb{T}_i^{pp} \qquad (i = 1, 2, \ldots, r) \\
&\qquad (\frac{\partial C_i}{\partial \boldsymbol{u}})^{\mathrm{T}} \boldsymbol{C}_{\boldsymbol{u}}^{\mathrm{T}} \boldsymbol{\mu}(t) + \int_{t_0}^{t} \hat{\boldsymbol{d}}_i^{L}(t,\sigma)\boldsymbol{\mu}(\sigma)\,\mathrm{d}\sigma + \int_{t}^{t_f} \hat{\boldsymbol{d}}_i^{R}(t,\sigma)\boldsymbol{\mu}(\sigma)\mathrm{d}\sigma + \hat{d}_i^{A}(t) + \hat{\boldsymbol{d}}_i^{\boldsymbol{\pi}}(t)\boldsymbol{\pi} = 0
\end{aligned}
\tag{70}
$$

and

$$
\begin{bmatrix} \boldsymbol{M}_{s1} \\ \boldsymbol{M}_{s2} \end{bmatrix} \boldsymbol{\pi} = -\begin{bmatrix} \boldsymbol{r}_{s1} \\ \boldsymbol{r}_{s2} \end{bmatrix} - \begin{bmatrix} \boldsymbol{g}_{\boldsymbol{x}_f} \int_{t_0}^{t_f} \hat{\boldsymbol{h}}_2(t)\boldsymbol{\mu}(t)\mathrm{d}t \\ \boldsymbol{0} \end{bmatrix}
\tag{71}
$$

where

$$
\hat{\boldsymbol{d}}_i^{L}(t,\sigma) = \left(\frac{\partial C_i}{\partial \boldsymbol{x}}(t)\right)^{\mathrm{T}} \left(\int_{t_0}^{\sigma} \boldsymbol{H}_o(t,s)\boldsymbol{H}_o^{\mathrm{T}}(\sigma,s)\,\mathrm{d}s\right) \boldsymbol{C}_{\boldsymbol{x}}^{\mathrm{T}}(\sigma) + \left(\frac{\partial C_i}{\partial \boldsymbol{x}}(t)\right)^{\mathrm{T}} \boldsymbol{H}_o(t,\sigma)\boldsymbol{C}_{\boldsymbol{u}}^{\mathrm{T}}(\sigma)
\tag{72}
$$

$$
\hat{\boldsymbol{d}}_i^{R}(t,\sigma) = \left(\frac{\partial C_i}{\partial \boldsymbol{x}}(t)\right)^{\mathrm{T}} \left(\int_{t_0}^{t} \boldsymbol{H}_o(t,s)\boldsymbol{H}_o^{\mathrm{T}}(\sigma,s)\,\mathrm{d}s\right) \boldsymbol{C}_{\boldsymbol{x}}^{\mathrm{T}}(\sigma) + \left(\frac{\partial C_i}{\partial \boldsymbol{u}}(t)\right)^{\mathrm{T}} \boldsymbol{H}_o^{\mathrm{T}}(\sigma,t)\boldsymbol{C}_{\boldsymbol{x}}^{\mathrm{T}}(\sigma)
\tag{73}
$$

$$
\hat{d}_i^{A}(t) = (\frac{\partial C_i}{\partial \boldsymbol{x}})^{\mathrm{T}} \int_{t_0}^{t} \boldsymbol{H}_o(t,s)\boldsymbol{p}_{\boldsymbol{u}}\,\mathrm{d}s + (\frac{\partial C_i}{\partial \boldsymbol{u}})^{\mathrm{T}} \boldsymbol{p}_{\boldsymbol{u}}
\tag{74}
$$

$$
\hat{\boldsymbol{d}}_i^{\boldsymbol{\pi}}(t) = \left( (\frac{\partial C_i}{\partial \boldsymbol{x}})^{\mathrm{T}} \int_{t_0}^{t} \boldsymbol{H}_o(t,s)\boldsymbol{H}_o^{\mathrm{T}}(t_f,s)\,\mathrm{d}s + (\frac{\partial C_i}{\partial \boldsymbol{u}})^{\mathrm{T}} \boldsymbol{H}_o^{\mathrm{T}}(t_f,t) \right) \boldsymbol{g}_{\boldsymbol{x}_f}^{\mathrm{T}}
\tag{75}
$$

The $q \times q$ dimensional matrixes $\boldsymbol{M}_{s1}$, $\boldsymbol{M}_{s2}$ and the $q$ dimensional vectors $\boldsymbol{r}_{s1}$, $\boldsymbol{r}_{s2}$ are

$$
\boldsymbol{M}_{s1} = \boldsymbol{g}_{\boldsymbol{x}_f} \left(\int_{t_0}^{t_f} \boldsymbol{H}_o(t_f,t)\boldsymbol{H}_o^{\mathrm{T}}(t_f,t)\,\mathrm{d}t\right) \boldsymbol{g}_{\boldsymbol{x}_f}^{\mathrm{T}}
\tag{76}
$$

$$
\boldsymbol{M}_{s2} = (\boldsymbol{g}_{\boldsymbol{x}_f}\boldsymbol{f} + \boldsymbol{g}_{t_f})(\boldsymbol{g}_{\boldsymbol{x}_f}\boldsymbol{f} + \boldsymbol{g}_{t_f})^{\mathrm{T}}\Big|_{t_f}
\tag{77}
$$

$$
\boldsymbol{r}_{s1} = \boldsymbol{g}_{\boldsymbol{x}_f} \int_{t_0}^{t_f} \boldsymbol{H}_o(t_f,t)\boldsymbol{p}_{\boldsymbol{u}}\,\mathrm{d}t
\tag{78}
$$

$$
\boldsymbol{r}_{s2} = (\boldsymbol{g}_{\boldsymbol{x}_f}\boldsymbol{f} + \boldsymbol{g}_{t_f})(\varphi_{t_f} + \varphi_{\boldsymbol{x}_f}^{\mathrm{T}}\boldsymbol{f} + L)\Big|_{t_f}
\tag{79}
$$

$$
\hat{\boldsymbol{h}}_2(t) = \left(\int_{t_0}^{t} \boldsymbol{H}_o(t_f,s)\boldsymbol{H}_o^{\mathrm{T}}(t,s)\,\mathrm{d}s\right) \boldsymbol{C}_{\boldsymbol{x}}^{\mathrm{T}} + \boldsymbol{H}_o(t_f,t)\boldsymbol{C}_{\boldsymbol{u}}^{\mathrm{T}}
\tag{80}
$$

**Proof**: Regarding the argument that $\mathbb{T}_i^{pp} = \mathbb{T}_i^{p}$ $(i = 1, 2, \ldots, r)$ for the optimal solution of Problem 1, this is because any time point $t$ in $\mathbb{T}_i^{p}$ also belongs to $\mathbb{T}_i^{pp}$ ultimately, or this activated inequality path constraint at time $t$ will become inactive.

For the optimal values of $\boldsymbol{\mu}(t)$ and $\boldsymbol{\pi}$, since $\boldsymbol{K}$ may be arbitrary right-dimensional positive-definite matrix, we set $\boldsymbol{K} = \boldsymbol{I}$ in Eq. (64), where $\boldsymbol{I}$ is the $m \times m$ dimensional identity matrix. Then we get Eq. (70) after appropriate simplification. From Eq. (61), again because $\boldsymbol{K}$ may be arbitrary right-dimensional positive-definite matrix and $k_{t_f}$ may be arbitrary positive constant, we consider three cases, that are, i) $\boldsymbol{K} = \boldsymbol{I}$, $k_{t_f} = 1$, ii) $\boldsymbol{K} = 2\boldsymbol{I}$, $k_{t_f} = 1$, and iii) $\boldsymbol{K} = \boldsymbol{I}$, $k_{t_f} = 2$. By comparing the three cases of substituting the specific values into Eq. (61), we may obtain Eq. (71), which is irrelevant to the specific value of $\boldsymbol{K}$ and $k_{t_f}$. ■

*C. Equivalence to the classic optimality conditions*

Actually, Eqs. (51) and (52) are the first-order optimality conditions for Problem 1 without the employment of costates. We will show that they are equivalent to the classic ones with costates [28]. By the direct adjoining method [11], we may construct the augmented functional as

$$\bar{J} = \varphi(\boldsymbol{x}(t_f), t_f) + \bar{\boldsymbol{\pi}}^{\mathrm{T}} \boldsymbol{g}\left(\boldsymbol{x}(t_f), t\right) + \int_{t_0}^{t_f} \left(L + \boldsymbol{\lambda}^{\mathrm{T}}(\boldsymbol{f} - \dot{\boldsymbol{x}}) + \bar{\boldsymbol{\mu}}^{\mathrm{T}} \boldsymbol{C}\right) \mathrm{d}t \tag{81}$$

and

$$\begin{aligned} &\bar{\mu}_i(t) C_i = 0 \quad when \quad t \in \mathbb{T}_i^p \qquad (i = 1, 2, ..., r) \\ &\bar{\mu}_i(t) = 0 \qquad when \quad t \notin \mathbb{T}_i^p \qquad (i = 1, 2, ..., r) \end{aligned} \tag{82}$$

where $\boldsymbol{\lambda} \in \mathbb{R}^n$ is the costate variable, $\bar{\boldsymbol{\pi}} \in \mathbb{R}^q$ is Lagrange multiplier parameter, and $\bar{\boldsymbol{\mu}}(t) \in \mathbb{R}^r$ is the KKT multiplier variable. Then the first-order variation may be derived as

$$\begin{aligned} \delta \bar{J} = &\left(\varphi_{t_f} + \bar{\boldsymbol{\pi}}^{\mathrm{T}} \boldsymbol{g}_{t_f} + \bar{H}\right)\Big|_{t_f} \delta t_f + \left(\boldsymbol{\lambda}(t_f) - \varphi_{\boldsymbol{x}_f} - \boldsymbol{g}_{\boldsymbol{x}_f}{}^{\mathrm{T}} \bar{\boldsymbol{\pi}}\right)^{\mathrm{T}} \delta \boldsymbol{x}(t_f) \\ &+ \int_{t_0}^{t_f} \left((\bar{H}_{\boldsymbol{\lambda}} - \dot{\boldsymbol{x}})^{\mathrm{T}} \delta \boldsymbol{\lambda} + (\bar{H}_{\boldsymbol{x}} + \dot{\boldsymbol{\lambda}})^{\mathrm{T}} \delta \boldsymbol{x} + \bar{H}_{\boldsymbol{u}}{}^{\mathrm{T}} \delta \boldsymbol{u} + \boldsymbol{C}^{\mathrm{T}} \delta \bar{\boldsymbol{\mu}}\right) \mathrm{d}t \end{aligned} \tag{83}$$

where $\bar{H} = L + \boldsymbol{\lambda}^{\mathrm{T}} \boldsymbol{f} + \bar{\boldsymbol{\mu}}^{\mathrm{T}} \boldsymbol{C}$ is the augmented Hamiltonian. Through $\delta \bar{J} = 0$, we have

$$\dot{\boldsymbol{\lambda}} + \bar{H}_{\boldsymbol{x}} = \dot{\boldsymbol{\lambda}} + L_{\boldsymbol{x}} + \boldsymbol{f}_{\boldsymbol{x}}{}^{\mathrm{T}} \boldsymbol{\lambda} + \boldsymbol{C}_{\boldsymbol{x}}{}^{\mathrm{T}} \bar{\boldsymbol{\mu}} = \boldsymbol{0} \tag{84}$$

$$\bar{H}_{\boldsymbol{u}} = L_{\boldsymbol{u}} + \boldsymbol{f}_{\boldsymbol{u}}{}^{\mathrm{T}} \boldsymbol{\lambda} + \boldsymbol{C}_{\boldsymbol{u}}{}^{\mathrm{T}} \bar{\boldsymbol{\mu}} = \boldsymbol{0} \tag{85}$$

and the transversality conditions

$$(L + \boldsymbol{\lambda}^{\mathrm{T}} \boldsymbol{f} + \varphi_{t_f} + \bar{\boldsymbol{\pi}}^{\mathrm{T}} \boldsymbol{g}_{t_f})\Big|_{t_f} = 0 \tag{86}$$

$$\boldsymbol{\lambda}(t_f) - \varphi_{\boldsymbol{x}_f} - \boldsymbol{g}_{\boldsymbol{x}_f}{}^{\mathrm{T}} \bar{\boldsymbol{\pi}} = \boldsymbol{0} \tag{87}$$

and the KKT condition (82).

**Theorem 4:** For Problem 1, the optimality conditions given by Eqs. (51) and (52) are equivalent to the optimality conditions given by Eqs. (82), (84)-(86).

**Proof**: Define a variable $\boldsymbol{\gamma}(t)$ as

$$\boldsymbol{\gamma}(t) = \boldsymbol{\Phi}_o{}^{\mathrm{T}}(t_f, t) \varphi_{\boldsymbol{x}_f} + \boldsymbol{\Phi}_o{}^{\mathrm{T}}(t_f, t) \boldsymbol{g}_{\boldsymbol{x}_f}{}^{\mathrm{T}} \boldsymbol{\pi} + \int_t^{t_f} \boldsymbol{\Phi}_o{}^{\mathrm{T}}(\sigma, t) L_{\boldsymbol{x}}(\sigma) \mathrm{d}\sigma + \int_t^{t_f} \boldsymbol{\Phi}_o{}^{\mathrm{T}}(\sigma, t) \boldsymbol{C}_{\boldsymbol{x}}{}^{\mathrm{T}}(\sigma) \boldsymbol{\mu}(\sigma) \mathrm{d}\sigma \tag{88}$$

Then Eq. (51) is simplified as

$$L_{\boldsymbol{u}} + \boldsymbol{f}_u{}^{\mathrm{T}} \boldsymbol{\gamma} + \boldsymbol{C}_{\boldsymbol{u}}{}^{\mathrm{T}} \boldsymbol{\mu} = \boldsymbol{0} \tag{89}$$

Obviously, when $t = t_f$, there is

$$\boldsymbol{\gamma}(t_f) = \varphi_{\boldsymbol{x}_f} + \boldsymbol{g}_{\boldsymbol{x}_f}{}^{\mathrm{T}} \boldsymbol{\pi} \tag{90}$$

Under Eqs. (38) and (39), Eq. (68) may be simplified as

$$\frac{\delta J}{\delta \tau} = \left(\varphi_{t_f} + \varphi_{\boldsymbol{x}_f}{}^{\mathrm{T}} \boldsymbol{f} + L + \boldsymbol{\pi}^{\mathrm{T}} (\boldsymbol{g}_{\boldsymbol{x}_f} \boldsymbol{f} + \boldsymbol{g}_{t_f})\right)\Big|_{t_f} \frac{\delta t_f}{\delta \tau} + \int_{t_0}^{t_f} \left(\boldsymbol{p}_{\boldsymbol{u}}^{pc}\right)^{\mathrm{T}} \frac{\delta \boldsymbol{u}}{\delta \tau} \mathrm{d}t \tag{91}$$

which holds in the feasible solution domain $\mathbb{D}_o$. Further combined with Eq. (90) and ignore $\delta \tau$, we have

$$\delta J=\left(\varphi_{t_f}+\boldsymbol{\pi}^{\mathrm{T}}\boldsymbol{g}_{t_f}+L+\boldsymbol{\gamma}(t_f)^{\mathrm{T}}\boldsymbol{f}\right)\Big|_{t_f}\delta t_f+\int_{t_0}^{t_f}\left(\boldsymbol{p}_{\boldsymbol{u}}^{pc}\right)^{\mathrm{T}}\delta\boldsymbol{u}\,\mathrm{d}t \tag{92}$$

Eq. (92) obviously holds at the optimal solution. Compare Eq. (92) with Eq. (83) and consider the variation of the terminal time. To achieve the extremal condition, Eqs. (52) and (86) should be same, i.e.

$$\left(L+\boldsymbol{\gamma}(t_f)^{\mathrm{T}}\boldsymbol{f}+\varphi_{t_f}+\boldsymbol{\pi}^{\mathrm{T}}\boldsymbol{g}_{t_f}\right)\Big|_{t_f}=\left(L+\boldsymbol{\lambda}^{\mathrm{T}}\boldsymbol{f}+\varphi_{t_f}+\overline{\boldsymbol{\pi}}^{\mathrm{T}}\boldsymbol{g}_{t_f}\right)\Big|_{t_f} \tag{93}$$

Since Eq. (93) generally hold for arbitrary $\boldsymbol{g}$ and $\boldsymbol{f}$ , we can conclude that

$$\boldsymbol{\pi}=\overline{\boldsymbol{\pi}} \tag{94}$$

$$\boldsymbol{\gamma}(t_f)=\boldsymbol{\lambda}(t_f) \tag{95}$$

Therefor Eq. (90) is same to Eq. (87). Consider the variation on the control within the feasible solution domain in Eqs. (92) and (83). There should be the conclusion that Eqs. (89) and (85) are identical, namely

$$L_{\boldsymbol{u}}+\boldsymbol{f}_{\boldsymbol{u}}{}^{\mathrm{T}}\boldsymbol{\lambda}+\boldsymbol{C}_{\boldsymbol{u}}{}^{\mathrm{T}}\overline{\boldsymbol{\mu}}=L_{\boldsymbol{u}}+\boldsymbol{f}_{\boldsymbol{u}}{}^{\mathrm{T}}\boldsymbol{\gamma}+\boldsymbol{C}_{\boldsymbol{u}}{}^{\mathrm{T}}\boldsymbol{\mu} \tag{96}$$

Because $\mathbb{T}_i^{pp}=\mathbb{T}_i^{p}\quad(i=1,2,...,r)$ for the optimal solution, Eq. (96) implies

$$\boldsymbol{\mu}(t)=\overline{\boldsymbol{\mu}}(t) \tag{97}$$

$$\boldsymbol{\gamma}(t)=\boldsymbol{\lambda}(t) \tag{98}$$

Differentiate $\boldsymbol{\gamma}(t)$ , as defined in Eq. (88), with respect to $t$ . In the process, we will use the Leibniz rule [29]

$$\frac{\mathrm{d}}{\mathrm{d}t}\left(\int_{b(t)}^{a(t)}h(\sigma,t)\,\mathrm{d}\sigma\right)=h\big(a(t),t\big)\frac{\mathrm{d}}{\mathrm{d}t}a(t)-h\big(b(t),t\big)\frac{\mathrm{d}}{\mathrm{d}t}b(t)+\int_{b(t)}^{a(t)}h_t(\sigma,t)\,\mathrm{d}\sigma \tag{99}$$

and the property of $\boldsymbol{\Phi}_o(\sigma,t)$ [26]

$$\frac{\partial\boldsymbol{\Phi}_o(\sigma,t)}{\partial t}=-\boldsymbol{\Phi}_o(\sigma,t)\boldsymbol{f}_{\boldsymbol{x}}(t) \tag{100}$$

$$\boldsymbol{\Phi}_o(t,t)=\boldsymbol{1} \tag{101}$$

where $\boldsymbol{1}$ is the $n\times n$ dimensional identity matrix. Then we have

$$\begin{aligned}\frac{\mathrm{d}}{\mathrm{d}t}\boldsymbol{\gamma}(t)&=-\boldsymbol{f}_{\boldsymbol{x}}{}^{\mathrm{T}}\boldsymbol{\Phi}_o{}^{\mathrm{T}}(t_f,t)\varphi_{\boldsymbol{x}_f}-\boldsymbol{f}_{\boldsymbol{x}}{}^{\mathrm{T}}\boldsymbol{\Phi}_o{}^{\mathrm{T}}(t_f,t)\boldsymbol{g}_{\boldsymbol{x}_f}{}^{\mathrm{T}}\boldsymbol{\pi}-L_{\boldsymbol{x}}-\boldsymbol{f}_{\boldsymbol{x}}{}^{\mathrm{T}}\int_t^{t_f}\boldsymbol{\Phi}_o{}^{\mathrm{T}}(\sigma,t)L_{\boldsymbol{x}}(\sigma)\,\mathrm{d}\sigma-\boldsymbol{C}_{\boldsymbol{x}}{}^{\mathrm{T}}(\sigma)\boldsymbol{\mu}(\sigma)-\boldsymbol{f}_{\boldsymbol{x}}{}^{\mathrm{T}}\int_t^{t_f}\boldsymbol{\Phi}_o{}^{\mathrm{T}}(\sigma,t)\boldsymbol{C}_{\boldsymbol{x}}{}^{\mathrm{T}}(\sigma)\boldsymbol{\mu}(\sigma)\mathrm{d}\sigma\\&=-L_{\boldsymbol{x}}-\boldsymbol{C}_{\boldsymbol{x}}{}^{\mathrm{T}}(\sigma)\boldsymbol{\mu}(\sigma)-\boldsymbol{f}_{\boldsymbol{x}}{}^{\mathrm{T}}\left(\boldsymbol{\Phi}_o{}^{\mathrm{T}}(t_f,t)\varphi_{\boldsymbol{x}_f}+\boldsymbol{\Phi}_o{}^{\mathrm{T}}(t_f,t)\boldsymbol{g}_{\boldsymbol{x}_f}{}^{\mathrm{T}}\boldsymbol{\pi}+\int_t^{t_f}\boldsymbol{\Phi}_o{}^{\mathrm{T}}(\sigma,t)L_{\boldsymbol{x}}(\sigma)\,\mathrm{d}\sigma+\int_t^{t_f}\boldsymbol{\Phi}_o{}^{\mathrm{T}}(\sigma,t)\boldsymbol{C}_{\boldsymbol{x}}{}^{\mathrm{T}}(\sigma)\boldsymbol{\mu}(\sigma)\mathrm{d}\sigma\right)\\&=-L_{\boldsymbol{x}}-\boldsymbol{f}_{\boldsymbol{x}}{}^{\mathrm{T}}\boldsymbol{\gamma}(t)-\boldsymbol{C}_{\boldsymbol{x}}{}^{\mathrm{T}}(\sigma)\boldsymbol{\mu}(\sigma)\end{aligned} \tag{102}$$

With Eq. (97), this means $\boldsymbol{\gamma}(t)$ conforms to the same dynamics as the costates $\boldsymbol{\lambda}(t)$ , and Eqs. (102) and (84) are exactly the same. Thus the theorem is proved. ■

From Theorems 3 and 4, we have got the explicit analytic relations of the costates $\boldsymbol{\lambda}$ , the Lagrange multiplier parameters $\overline{\boldsymbol{\pi}}$ and the KKT multiplier variables $\overline{\boldsymbol{\mu}}$ for the classic treatment in Eq. (81) to the original (state and control) variables, which formerly can only be obtained numerically by solving the BVP. To present these results more clearly, they are summarized in Table 1.

**Table 1** The classic optimality conditions and the costate-free optimality conditions for Problem 1

| Feasibility conditions | i) $\dot{\boldsymbol{x}} = \boldsymbol{f}(\boldsymbol{x},\boldsymbol{u},t)$ ; ii) $\boldsymbol{x}(t_0) = \boldsymbol{x}_0$ ; iii) $\boldsymbol{g}\left(\boldsymbol{x}(t_f),t_f\right) = \boldsymbol{0}$ ; iv) $\boldsymbol{C}\left(\boldsymbol{x}(t),\boldsymbol{u}(t),t\right) \leq \boldsymbol{0}$ . |
| --- | --- |
| Classic optimality conditions | i) $\dot{\boldsymbol{\lambda}} + L_{\boldsymbol{x}} + \boldsymbol{f_x}^{\mathrm{T}}\boldsymbol{\lambda} + \boldsymbol{C_x}^{\mathrm{T}}\overline{\boldsymbol{\mu}} = \boldsymbol{0}$ ; ii) $L_{\boldsymbol{u}} + \boldsymbol{f_u}^{\mathrm{T}}\boldsymbol{\lambda} + \boldsymbol{C_u}^{\mathrm{T}}\overline{\boldsymbol{\mu}} = \boldsymbol{0}$ ; iii) $(L + \boldsymbol{\lambda}^{\mathrm{T}}\boldsymbol{f} + \varphi_{t_f} + \overline{\boldsymbol{\pi}}^{\mathrm{T}}\boldsymbol{g}_{t_f})\Big\|_{t_f} = 0$ ;<br>iv) $\boldsymbol{\lambda}(t_f) - \varphi_{\boldsymbol{x}_f} - \overline{\boldsymbol{\pi}}^{\mathrm{T}}\boldsymbol{g}_{\boldsymbol{x}_f} = \boldsymbol{0}$ ; v) $\begin{array}{llll} \overline{\mu}_i(t)C_i = 0 & when & t \in \mathbb{T}_i^p & (i = 1,2,...,r) \\ \overline{\mu}_i(t) = 0 & when & t \notin \mathbb{T}_i^p & (i = 1,2,...,r) \end{array}$ . |
| Costate-free optimality conditions | i) $\boldsymbol{p}_{\boldsymbol{u}}^{pc} = \boldsymbol{0}$ in which $\boldsymbol{\pi}$ is calculated by Eq. (41) and $\boldsymbol{\mu}(t)$ is determined by Eq. (46);<br>ii) $L + \varphi_{t_f} + \varphi_{\boldsymbol{x}_f}{}^{\mathrm{T}}\boldsymbol{f} + \boldsymbol{\pi}^{\mathrm{T}}(\boldsymbol{g}_{\boldsymbol{x}_f}\boldsymbol{f} + \boldsymbol{g}_{t_f})\Big\|_{t_f} = 0$ where $\boldsymbol{\pi}$ is calculated by Eq. (41). |
| Analytic relations | i) $\boldsymbol{\lambda}(t) = \boldsymbol{\Phi}_o{}^{\mathrm{T}}(t_f,t)\left(\varphi_{\boldsymbol{x}_f} + \boldsymbol{g}_{\boldsymbol{x}_f}{}^{\mathrm{T}}\boldsymbol{\pi}\right) + \int_t^{t_f}\boldsymbol{\Phi}_o{}^{\mathrm{T}}(\sigma,t)\left(L_{\boldsymbol{x}}(\sigma) + \boldsymbol{C_x}^{\mathrm{T}}(\sigma)\boldsymbol{\mu}(\sigma)\right)\mathrm{d}\sigma$ ;<br>ii) $\overline{\boldsymbol{\pi}} = -\boldsymbol{M}^{-1}\boldsymbol{r}$ where $\boldsymbol{M}$ is given in Eq. (42) and $\boldsymbol{r}$ is given in Eq. (43);<br>iii) $(\frac{\partial C_i}{\partial \boldsymbol{u}})^{\mathrm{T}}\boldsymbol{K}\boldsymbol{C_u}^{\mathrm{T}}\overline{\boldsymbol{\mu}}(t) + \int_{t_0}^{t_f}\boldsymbol{d}_i^W(t,\sigma)\overline{\boldsymbol{\mu}}(\sigma)\mathrm{d}\sigma + \int_{t_0}^{t}\boldsymbol{d}_i^L(t,\sigma)\overline{\boldsymbol{\mu}}(\sigma)\mathrm{d}\sigma + \int_t^{t_f}\boldsymbol{d}_i^R(t,\sigma)\overline{\boldsymbol{\mu}}(\sigma)\mathrm{d}\sigma + d_i^A(t) = 0 \quad for \;\; \overline{\mu}_i(t) \neq 0$<br>with $\boldsymbol{d}_i^W(t,\sigma)$ , $\boldsymbol{d}_i^L(t,\sigma)$ , $\boldsymbol{d}_i^R(t,\sigma)$ and $d_i^A(t)$ defined in Eqs. (47)-(50). |

After the proof of Theorem 4, now the variables evolving direction using the VEM is easy to determine and the optimal solution of Problem 1 will be sought with theoretical guarantee.

**Theorem 5:** Solving the IVP with respect to $\tau$ , defined by the variation dynamic evolution equations (25), (38) and (39) from a feasible initial solution, when $\tau \to +\infty$ , $(\boldsymbol{x},\boldsymbol{u})$ will satisfy the optimality conditions of Problem 1.

**Proof**: By Lemma 2 and with Eq. (17) as the Lyapunov functional, we may claim that the minimum solution of Problem 1 is an asymptotically stable solution within the feasibility domain $\mathbb{D}_o$ for the infinite-dimensional dynamics governed by Eqs. (25), (38) and (39). From a feasible initial solution, any evolution under these dynamics maintains the feasibility of the variables, and they also guarantee $\frac{\delta J}{\delta \tau} \leq 0$ . The functional $J$ will decrease until $\frac{\delta J}{\delta \tau} = 0$ , which occurs when $\tau \to +\infty$ due to the asymptotical approach. When $\frac{\delta J}{\delta \tau}$ =0, this determines the optimal conditions, namely, Eqs. (51) and (52). ■

*D. Formulation of EPDE*

Use the partial differential operator “ $\partial$ ” and the differential operator “ d ” to reformulate the variation dynamic evolution equations, we may get the EPDE and the EDE as

$$\frac{\partial}{\partial \tau}\begin{bmatrix} \boldsymbol{x}(t,\tau) \\ \boldsymbol{u}(t,\tau) \end{bmatrix} = \begin{bmatrix} \int_{t_0}^{t}\boldsymbol{H}_o(t,s)\frac{\partial \boldsymbol{u}(s,\tau)}{\partial \tau}\mathrm{d}s \\ -\boldsymbol{K}\boldsymbol{p}_{\boldsymbol{u}}^{pc} \end{bmatrix} \tag{103}$$

$$\frac{\mathrm{d}t_f}{\mathrm{d}\tau} = -k_{t_f}\left(L + \varphi_{t_f} + \varphi_{\boldsymbol{x}_f}{}^{\mathrm{T}}\boldsymbol{f} + \boldsymbol{\pi}^{\mathrm{T}}(\boldsymbol{g}_{\boldsymbol{x}_f}\boldsymbol{f} + \boldsymbol{g}_{t_f})\right)\Big|_{t_f} \tag{104}$$

Put into this perspective, the definite conditions are the initial guess of $t_f$ , i.e., $t_f\big|_{\tau=0} = \tilde{t}_f$ , and

$$\begin{bmatrix} \boldsymbol{x}(t,\tau) \\ \boldsymbol{u}(t,\tau) \end{bmatrix}\Bigg|_{\tau=0} = \begin{bmatrix} \tilde{\boldsymbol{x}}(t) \\ \tilde{\boldsymbol{u}}(t) \end{bmatrix} \tag{105}$$

where $\tilde{\boldsymbol{x}}(t)$ and $\tilde{\boldsymbol{u}}(t)$ are the feasible initial solutions.

Eqs. (103) and (104) realize the anticipated variable evolving along the variation time $\tau$ as depicted in Fig. 1. The initial conditions of $\boldsymbol{x}(t,\tau)$ and $\boldsymbol{u}(t,\tau)$ at $\tau = 0$ belong to the feasible solution domain and their values at $\tau = +\infty$ represent the optimal solution of the OCP. The right part of the EPDE (103) is also only a vector function of time $t$. Thus we may apply the semi-discrete method to discretize it along the normal time dimension and further use ODE integration methods to get the numerical solution. Meanwhile, the Lagrange multiplier parameters $\boldsymbol{\pi}$ and the KKT multiplier variables $\boldsymbol{\mu}(t)$ need to be solved during the evolution process. To determine the right functions of Eqs. (103) and (104) at any variation time $\tau$, generally we compute $\boldsymbol{\mu}(t)$ through Eq. (46) first, and then use Eq. (41) to get $\boldsymbol{\pi}$. For the integral equation in Eq. (46), it may be solved numerically at the discretization time points that belong to $\mathbb{T}_i^{pp}$ $(i = 1,2,...,r)$. Note that whether a specific time point belongs to $\mathbb{T}_i^{pp}$ $(i = 1,2,...,r)$ may be determined in light of Theorems 1 and 2.

## IV. Discussion

### *A. Various path constraints*

The path constraints (21) defined in Problem 1 take the mixed state-control form. Actually they can also include other parallel forms. For example, if the path constraint is in the form of pure-state type, i.e.

$$\boldsymbol{C}\left(\boldsymbol{x}(t),t\right) \le \boldsymbol{0} \tag{106}$$

then $\boldsymbol{C}_{\boldsymbol{u}} = \boldsymbol{0}$ and all terms in the evolution equations relevant to $\boldsymbol{C}_{\boldsymbol{u}}$ vanish. When the path constraint takes the pure-control form as

$$\boldsymbol{C}\left(\boldsymbol{u}(t),t\right) \le \boldsymbol{0} \tag{107}$$

then $\boldsymbol{C}_{\boldsymbol{x}} = \boldsymbol{0}$ and the terms relevant to $\boldsymbol{C}_{\boldsymbol{x}}$ disappear in the evolution equations. In particular, the integral equation in Eq. (46) that determines $\boldsymbol{\mu}(t)$ is now simplified as

$$(\frac{\partial C_i}{\partial \boldsymbol{u}})^{\mathrm{T}} \boldsymbol{K}\boldsymbol{C}_{\boldsymbol{u}}^{\mathrm{T}}\boldsymbol{\mu}(t) + \int_{t_0}^{t_f} \boldsymbol{d}_i^W(t,\sigma)\boldsymbol{\mu}(\sigma)\mathrm{d}\sigma + d_i^A(t) = 0 \tag{108}$$

which is a typical Fredholm integral equation.

Besides the inequality type path constraint, we may also investigate the equality type. Consider an extreme case that all the path constraints in (21) take the equality form, such as

$$\boldsymbol{C}\left(\boldsymbol{x}(t),\boldsymbol{u}(t),t\right) = \boldsymbol{0} \tag{109}$$

For such case, the treatment in deriving the evolution equations is still same, while now the determination of the multiplier variable $\boldsymbol{\mu}$ becomes simpler, just through solving the following integral equation for the whole time horizon $[t_0, t_f]$ as

$$\boldsymbol{C}_{\boldsymbol{u}}\boldsymbol{K}\boldsymbol{C}_{\boldsymbol{u}}^{\mathrm{T}}\boldsymbol{\mu}(\mathrm{t}) + \int_{t_0}^{t_f} \boldsymbol{D}^W(t,\sigma)\boldsymbol{\mu}(\sigma)\mathrm{d}\sigma + \int_{t_0}^{t} \boldsymbol{D}^L(t,\sigma)\boldsymbol{\mu}(\sigma)\,\mathrm{d}\,\sigma + \int_{t}^{t_f} \boldsymbol{D}^R(t,\sigma)\boldsymbol{\mu}(\sigma)\mathrm{d}\sigma + \boldsymbol{d}^A = \boldsymbol{0} \tag{110}$$

where

$$\boldsymbol{D}^W(t,\sigma) = -\boldsymbol{C}_{\boldsymbol{x}}(t)\left(\int_{t_0}^{t} \boldsymbol{H}_o(t,s)\boldsymbol{K}\boldsymbol{H}_o^{\mathrm{T}}(t_f,s)\,\mathrm{d}\,s\right)\boldsymbol{g}_{\boldsymbol{x}_f}^{\mathrm{T}}\boldsymbol{M}^{-1}\boldsymbol{g}_{\boldsymbol{x}_f}\boldsymbol{h}_2(\sigma) - \boldsymbol{C}_{\boldsymbol{u}}(t)\boldsymbol{K}\boldsymbol{H}_o^{\mathrm{T}}(t_f,t)\boldsymbol{g}_{\boldsymbol{x}_f}^{\mathrm{T}}\boldsymbol{M}^{-1}\boldsymbol{g}_{\boldsymbol{x}_f}\boldsymbol{h}_2(\sigma) \tag{111}$$

$$\boldsymbol{D}^{L}(t,\sigma)=\boldsymbol{C}_{\boldsymbol{x}}(t)\left(\int_{t_0}^{\sigma}\boldsymbol{H}_o(t,s)\boldsymbol{K}\boldsymbol{H}_o^{\mathrm{T}}(\sigma,s)\,\mathrm{d}s\right)\boldsymbol{C}_{\boldsymbol{x}}^{\mathrm{T}}(\sigma)+\boldsymbol{C}_{\boldsymbol{x}}(t)\boldsymbol{H}_o(t,\sigma)\boldsymbol{K}\boldsymbol{C}_{\boldsymbol{u}}^{\mathrm{T}}(\sigma) \tag{112}$$

$$\boldsymbol{D}^{R}(t,\sigma)=\boldsymbol{C}_{\boldsymbol{x}}(t)\left(\int_{t_0}^{t}\boldsymbol{H}_o(t,s)\boldsymbol{K}\boldsymbol{H}_o^{\mathrm{T}}(\sigma,s)\,\mathrm{d}s\right)\boldsymbol{C}_{\boldsymbol{x}}^{\mathrm{T}}(\sigma)+\boldsymbol{C}_{\boldsymbol{u}}(t)\boldsymbol{K}\boldsymbol{H}_o^{\mathrm{T}}(\sigma,t)\boldsymbol{C}_{\boldsymbol{x}}^{\mathrm{T}}(\sigma) \tag{113}$$

$$\boldsymbol{d}^{A}(t)=\boldsymbol{C}_{\boldsymbol{x}}\int_{t_0}^{t}\boldsymbol{H}_o(t,s)\boldsymbol{K}\boldsymbol{p}_{\boldsymbol{u}}\,\mathrm{d}s-\boldsymbol{C}_{\boldsymbol{x}}\left(\int_{t_0}^{t}\boldsymbol{H}_o(t,s)\boldsymbol{K}\boldsymbol{H}_o^{\mathrm{T}}(t_f,s)\,\mathrm{d}s\right)\boldsymbol{g}_{\boldsymbol{x}_f}^{\mathrm{T}}\boldsymbol{M}^{-1}\boldsymbol{h}_1+\boldsymbol{C}_{\boldsymbol{u}}\boldsymbol{K}\boldsymbol{p}_{\boldsymbol{u}}-\boldsymbol{C}_{\boldsymbol{u}}\boldsymbol{K}\boldsymbol{H}_o^{\mathrm{T}}(t_f,t)\boldsymbol{g}_{\boldsymbol{x}_f}^{\mathrm{T}}\boldsymbol{M}^{-1}\boldsymbol{h}_1 \tag{114}$$

When there are both inequality and equality path constraints in the OCP, then Eqs. (46) and (110) need to be synthesized in seeking the multiplier variables.

*B. Simpler OCP formulations*

Since being derived for the general constrained OCPs, the equations obtained in this paper are generally applicable. Now we consider simpler OCP formulations.

If the terminal time $t_f$ is fixed in the OCP, then Eqs. (103) and (104) may be directly applied by setting $k_{t_f}=0$.

If there is no path constraint (21) in the OCP, then we do not need to solve $\boldsymbol{\mu}(t)$ and it may be set as $\boldsymbol{\mu}(t)=\boldsymbol{0}$ in Eqs. (103) and (104) to get the optimal solution.

When there is no terminal constraint (20), then we may let $\boldsymbol{\pi}=\boldsymbol{0}$ in Eqs. (103) and (104), and the KKT multiplier variables $\boldsymbol{\mu}(t)$ is now determined by

$$\begin{aligned}
&when\quad t\notin\mathbb{T}_i^{pp}\qquad (i=1,2,...,r)\\
&\qquad \mu_i(t)=0\\
&when\quad t\in\mathbb{T}_i^{pp}\qquad (i=1,2,...,r)\\
&\qquad \left(\frac{\partial C_i}{\partial \boldsymbol{u}}\right)^{\mathrm{T}}\boldsymbol{K}\boldsymbol{C}_{\boldsymbol{u}}^{\mathrm{T}}\boldsymbol{\mu}(t)+\int_{t_0}^{t}\boldsymbol{d}_i^{L}(t,\sigma)\boldsymbol{\mu}(\sigma)\,\mathrm{d}\sigma+\int_{t}^{t_f}\boldsymbol{d}_i^{R}(t,\sigma)\boldsymbol{\mu}(\sigma)\mathrm{d}\sigma+d_i^{A}(t)=0
\end{aligned} \tag{115}$$

with $\boldsymbol{d}_i^{L}(t,\sigma)$ defined in Eq. (48), $\boldsymbol{d}_i^{R}(t,\sigma)$ defined in Eq. (49), and $d_i^{A}(t)$ being

$$d_i^{A}(t)=\left(\frac{\partial C_i}{\partial \boldsymbol{x}}\right)^{\mathrm{T}}\int_{t_0}^{t}\boldsymbol{H}_o(t,s)\boldsymbol{K}\boldsymbol{p}_{\boldsymbol{u}}\,\mathrm{d}s+\left(\frac{\partial C_i}{\partial \boldsymbol{u}}\right)^{\mathrm{T}}\boldsymbol{K}\boldsymbol{p}_{\boldsymbol{u}} \tag{116}$$

*C. Generation of a feasible initial solution*

The derivation in Sec. III starts from the premise that the initial solutions $\tilde{\boldsymbol{x}}(t)$ and $\tilde{\boldsymbol{u}}(t)$ are feasible, namely, satisfying Eqs. (18) -(21). For the general constrained OCPs, usually it is not an easy task to find a feasible solution. We may either further develop the VEM to be valid in the infeasible solution domain as Refs. [20] and [21] did, or we find an alternative to determine a feasible initial solution. Here we choose the second way. Upon the work in Ref. [21], which solves the OCPs with arbitrary initialization, we can generate a feasible initial solution by solving the following Feasible Solution Searching Optimization Problem (FSSOP) as

**FSSOP**:

$$\begin{aligned}
&\min\quad J_{fs}=\int_{t_0}^{t_f}\left\{\sum_{i=1}^{r}w_i C_i\left(\boldsymbol{x}(t),\boldsymbol{u}(t),t\right)\right\}\mathrm{d}t\\
&s.t.\\
&\dot{\boldsymbol{x}}=\boldsymbol{f}(\boldsymbol{x},\boldsymbol{u},t)\\
&\boldsymbol{x}(t_0)=\boldsymbol{x}_0\\
&\boldsymbol{g}\left(\boldsymbol{x}(t_f),t_f\right)=\boldsymbol{0}
\end{aligned} \tag{117}$$

where $w_i$ is the weight coefficient for the $i$ th path constraint $C_i$ in (21). The dynamics constraint and the boundary conditions are same to those in Problem 1, while the terminal time $t_f$ may be different.

Through solving this FSSOP with the method in Ref. [21], we may obtain a solution that is feasible for Problem 1.

*D. Numerical soft barrier*

Theoretically, the evolution equations will precisely seek the optimal solution. During the variable evolution process, once the inequality path constraint at a specific time point is activated, the corresponding variation constraint (in Eq. (32)) will be triggered immediately to maintain the feasibility of solutions. However, since we resort to the numerical method for the solution, concretely by using the ODE integration methods to solve the transformed finite-dimensional IVPs, the numerical error is unavoidable, and this may lead to the violation of the path constraints. Refer to the strategy to eliminate the violations on the terminal IECs [19], the numerical soft barrier technique is again employed to remove the possible numerical error on the path constraints, by adapting the FPEOP (34) as

**Adapted FPEOP**:

$$\begin{aligned}
&\min \quad J_{t3} = \frac{1}{2} J_{t1} + \frac{1}{2} J_{t2} \\
&s.t. \\
&\frac{\delta \boldsymbol{g}}{\delta \tau} = \boldsymbol{0} \\
&\frac{\delta C_i}{\delta \tau} + k_C C_i = 0 \qquad t \in \mathbb{T}_i^{pp},\ i = 1, 2, ..., r
\end{aligned} \tag{118}$$

where $k_C$ is a positive constant and now the time set $\mathbb{T}_i^{pp}$ is defined as

$$\mathbb{T}_i^{pp} = \{ t \mid C_i(\boldsymbol{x}, \boldsymbol{u}, t) \geq 0, \frac{\delta C_i}{\delta \tau} + k_C C_i \leq 0 \text{ is an active IEC, } t \in [t_0, t_f] \} \tag{119}$$

Through solving the adapted FPEOP, the evolution equations derived are still similar except $d_i^A(t)$ in Eq. (46) is modified as

$$d_i^A(t) = (\frac{\partial C_i}{\partial \boldsymbol{x}})^{\mathrm{T}} \int_{t_0}^{t} \boldsymbol{H}_o(t,s) \boldsymbol{K} \boldsymbol{p}_{\boldsymbol{u}} \,\mathrm{d}s - (\frac{\partial C_i}{\partial \boldsymbol{x}})^{\mathrm{T}} \left( \int_{t_0}^{t} \boldsymbol{H}_o(t,s) \boldsymbol{K} \boldsymbol{H}_o^{\mathrm{T}}(t_f, s) \,\mathrm{d}s \right) \boldsymbol{g}_{\boldsymbol{x}_f}{}^{\mathrm{T}} \boldsymbol{M}^{-1} \boldsymbol{h}_1 + (\frac{\partial C_i}{\partial \boldsymbol{u}})^{\mathrm{T}} \boldsymbol{K} \left( \boldsymbol{p}_{\boldsymbol{u}} \right) - (\frac{\partial C_i}{\partial \boldsymbol{u}})^{\mathrm{T}} \boldsymbol{K} \boldsymbol{H}_o^{\mathrm{T}}(t_f, t) \boldsymbol{g}_{\boldsymbol{x}_f}{}^{\mathrm{T}} \boldsymbol{M}^{-1} \boldsymbol{h}_1 - k_C C_i \tag{120}$$

In this way, the possible violations on the path constraints due to the numerical error will be eliminated gradually.

## V. Illustrative Examples

First a linear example taken from Zhang [30] is solved.

**Example 1:** Consider the following dynamic system

$$\dot{\boldsymbol{x}} = \boldsymbol{A}\boldsymbol{x} + \boldsymbol{b}u$$

where $\boldsymbol{x} = \begin{bmatrix} x_1 \\ x_2 \end{bmatrix}$, $\boldsymbol{A} = \begin{bmatrix} 0 & 1 \\ 0 & 0 \end{bmatrix}$, and $\boldsymbol{b} = \begin{bmatrix} 0 \\ 1 \end{bmatrix}$. Find the solution that minimizes the performance index

$$J = t_f$$

with the control constraint

$$-1 \leq u \leq 1$$

and the boundary conditions

$$\boldsymbol{x}(t_0) = \begin{bmatrix} 1 \\ 1 \end{bmatrix}, \boldsymbol{x}(t_f) = \begin{bmatrix} 0 \\ 0 \end{bmatrix}$$

where the initial time $t_0 = 0$ is fixed.

In solving this example using the VEM, the path constraint is reformulated as

$$u^2 - 1 \le 0$$

Then the EPDE derived is

$$\frac{\partial}{\partial \tau}\begin{bmatrix} \boldsymbol{x} \\ u \end{bmatrix} = \begin{bmatrix} \int_{t_0}^{t} e^{\boldsymbol{A}(t-s)} \boldsymbol{b} \frac{\partial \boldsymbol{u}}{\partial \tau}(s)\, \mathrm{d}\, s \\ -K\left\{ \boldsymbol{b}^{\mathrm{T}} \left( e^{\boldsymbol{A}(t_f - t)} \right)^{\mathrm{T}} \boldsymbol{\pi} + 2u\mu(t) \right\} \end{bmatrix}$$

where $\boldsymbol{\pi}$ is solved by Eq. (41) and the scalar KKT multiplier variable $\mu(t)$ is determined by Eq. (46). The one-dimensional gain matrix $K$ was $K = 0.2$ and the scalar $k_{t_f}$ was $k_{t_f} = 0.1$. The barrier parameter $k_C$ in Eq. (120) was set to be 0.1. The definite conditions of the EPDE, i.e., the feasible initial guess of the states $\tilde{\boldsymbol{x}}(t)$ and the control $\tilde{u}(t)$, were obtained by solving the following FSSOP as

$$\begin{aligned}
&\min \quad J_{fs} = \frac{1}{2}\int_{t_0}^{t_f} u^2 \mathrm{d}t \\
&s.t. \\
&\dot{\boldsymbol{x}} = \boldsymbol{A}\boldsymbol{x} + \boldsymbol{b}u \\
&\boldsymbol{x}(t_0) = \begin{bmatrix} 1 \\ 1 \end{bmatrix}, \boldsymbol{x}(t_f) = \begin{bmatrix} 0 \\ 0 \end{bmatrix} \\
&t_0 = 0, t_f = 8
\end{aligned}$$

Note that $t_f$ =8s is also the initial guess of the terminal time $\tilde{t}_f$ for this example. Using the semi-discrete method, the time horizon $[t_0, t_f]$ was discretized uniformly with 41 points. Thus, a dynamic system with 124 states (including the terminal time) was obtained and the OCP was transformed to a finite-dimensional IVP. The ODE integrator "ode45" in Matlab, with default relative error tolerance $1\times10^{-3}$ and default absolute error tolerance $1\times10^{-6}$, was employed to solve the IVP. For comparison, the analytic solution by solving the BVP is also presented.

$$t \in [0,\ 1 + (\sqrt{6}/2)) \qquad\qquad t \in [1 + (\sqrt{6}/2),\ 1 + \sqrt{6}]$$

$$\begin{cases} \hat{x}_1 = -0.5t^2 + t + 1 \\ \hat{x}_2 = -t + 1 \\ \hat{\lambda}_1 = \sqrt{6}/3 \\ \hat{\lambda}_2 = -(\sqrt{6}/3)t + (\sqrt{6}/3) + 1 \\ \hat{u} = -1 \end{cases} \qquad \begin{cases} \hat{x}_1 = 0.5t^2 - (1+\sqrt{6})t + 3.5 + \sqrt{6} \\ \hat{x}_2 = t - 1 - \sqrt{6} \\ \hat{\lambda}_1 = \sqrt{6}/3 \\ \hat{\lambda}_2 = -(\sqrt{6}/3)t + (\sqrt{6}/3) + 1 \\ \hat{u} = 1 \end{cases}$$

Figs. 2 and 3 present the evolution process of $x_1(t)$ and $u(t)$ towards the analytic solutions, showing the asymptotically approach of the numerical results to the optimal. At $\tau$ = 300s, they are very close to the analytic solutions, and this demonstrates the effectiveness of the VEM. For the control results plotted in Fig. 3, it is shown that the control switch is accurately captured from the close-up. In Fig. 4, the states results are again compared with the analytic solution in the state plane, illustrating the evolution process of the states from a different angle. The profile for the terminal time is given in Fig 5. It monotonously decreases from $\tilde{t}_f$ = 8s and is almost unchanged after $\tau$ = 100s. At $\tau$ = 300s, we compute that $t_f$ = 3.44s, very close to the analytic result. Regarding the Lagrange multipliers that adjoin the terminal constraint, we computed that $\boldsymbol{\pi}=\begin{bmatrix}0.83\\-1.00\end{bmatrix}$ at $\tau$ =300s. From the analytic relation to the costates in Table 1, we have

$$\boldsymbol{\lambda}(t)=\boldsymbol{\Phi}_o^{\mathrm{T}}(t_f,t)\boldsymbol{\pi}=\left(e^{A(t_f-t)}\right)^{\mathrm{T}}\boldsymbol{\pi}=\begin{bmatrix}1 & 0\\ t_f-t & 1\end{bmatrix}\begin{bmatrix}0.83\\-1.00\end{bmatrix}=\begin{bmatrix}0.83\\-0.83t+1.86\end{bmatrix}$$

This is close to the analytic solution of $\begin{bmatrix}\hat{\lambda}_1\\ \hat{\lambda}_2\end{bmatrix}$. In Fig. 6, the numerical solution of KKT multiplier variable $\mu(t)$ at $\tau$ = 300s is presented, and the sharp angle of the curve at the control switch time point is clearly shown.

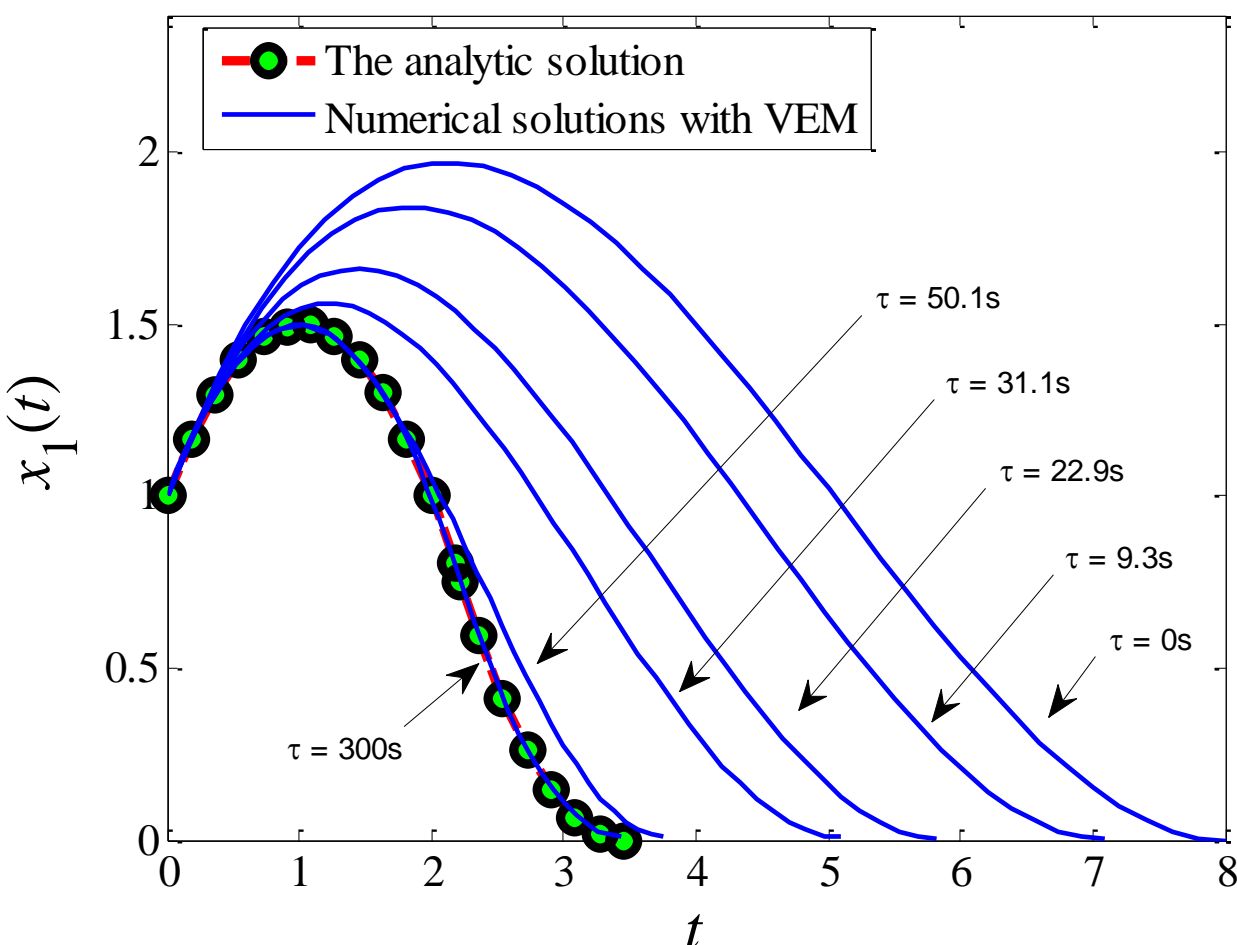

Fig. 2 The evolution of numerical solutions of $x_1$ to the analytic solution.

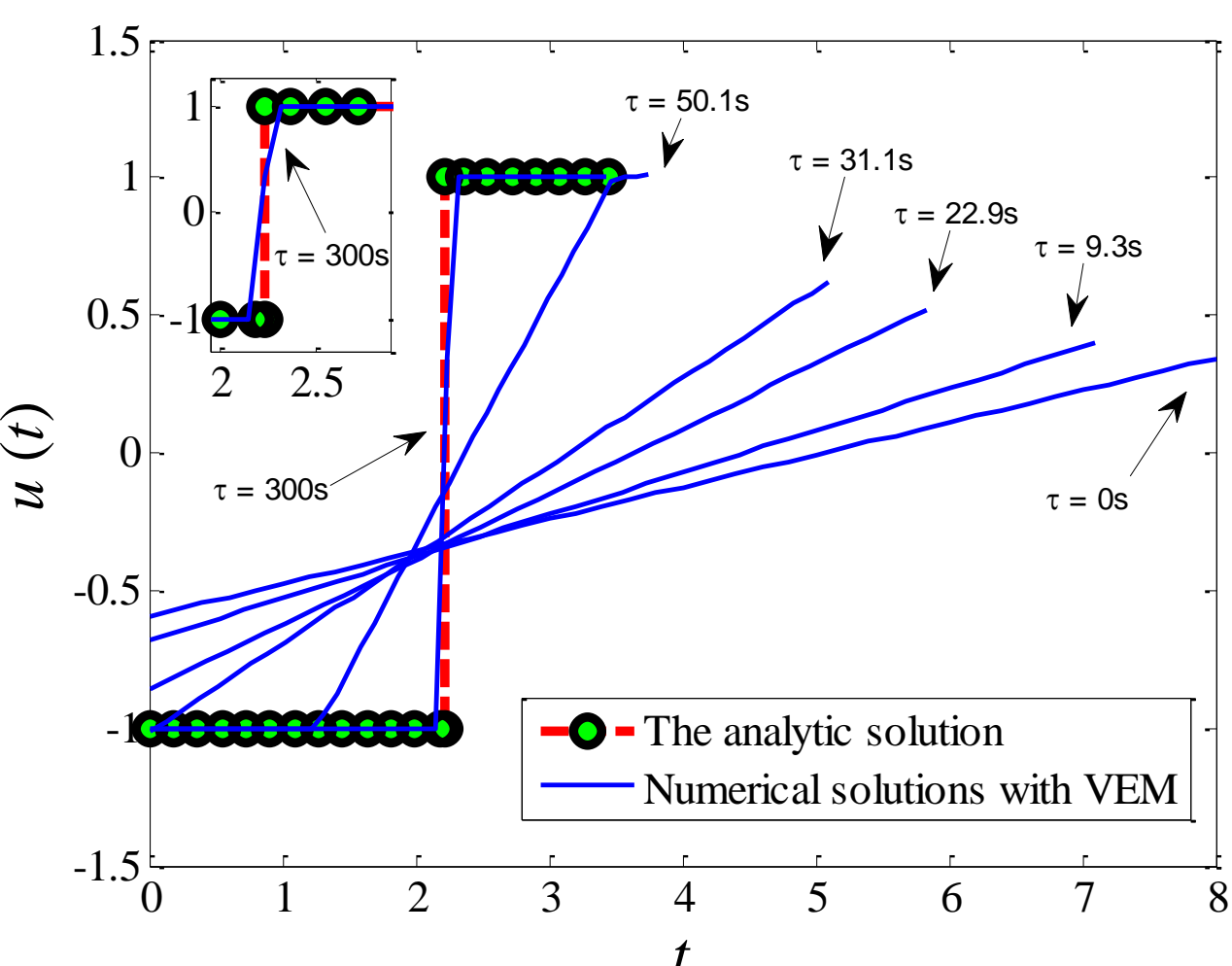

Fig. 3 The evolution of numerical solutions of $u$ to the analytic solution.

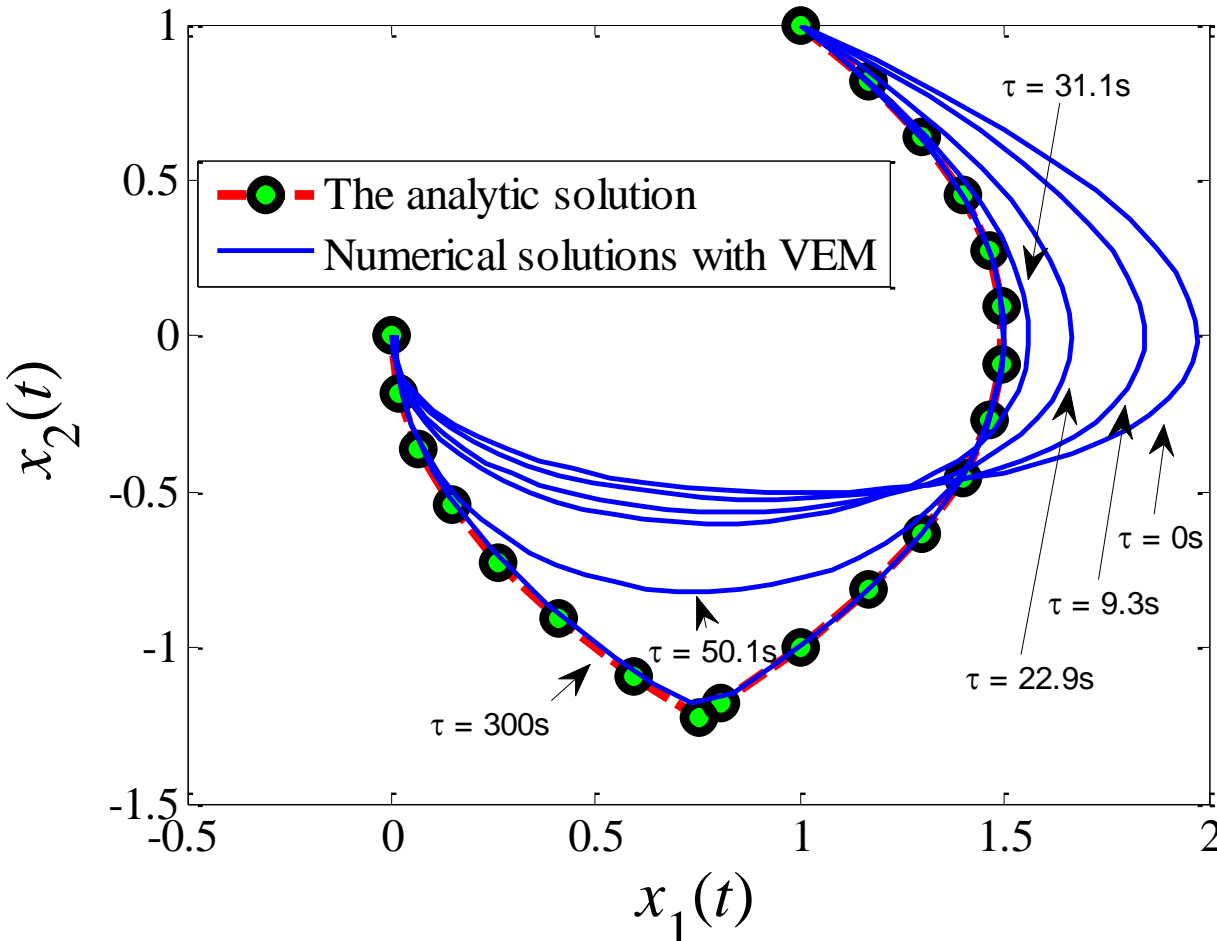

Fig. 4 The evolution of numerical solutions in $x_1 x_2$ state plane to the analytic solution.

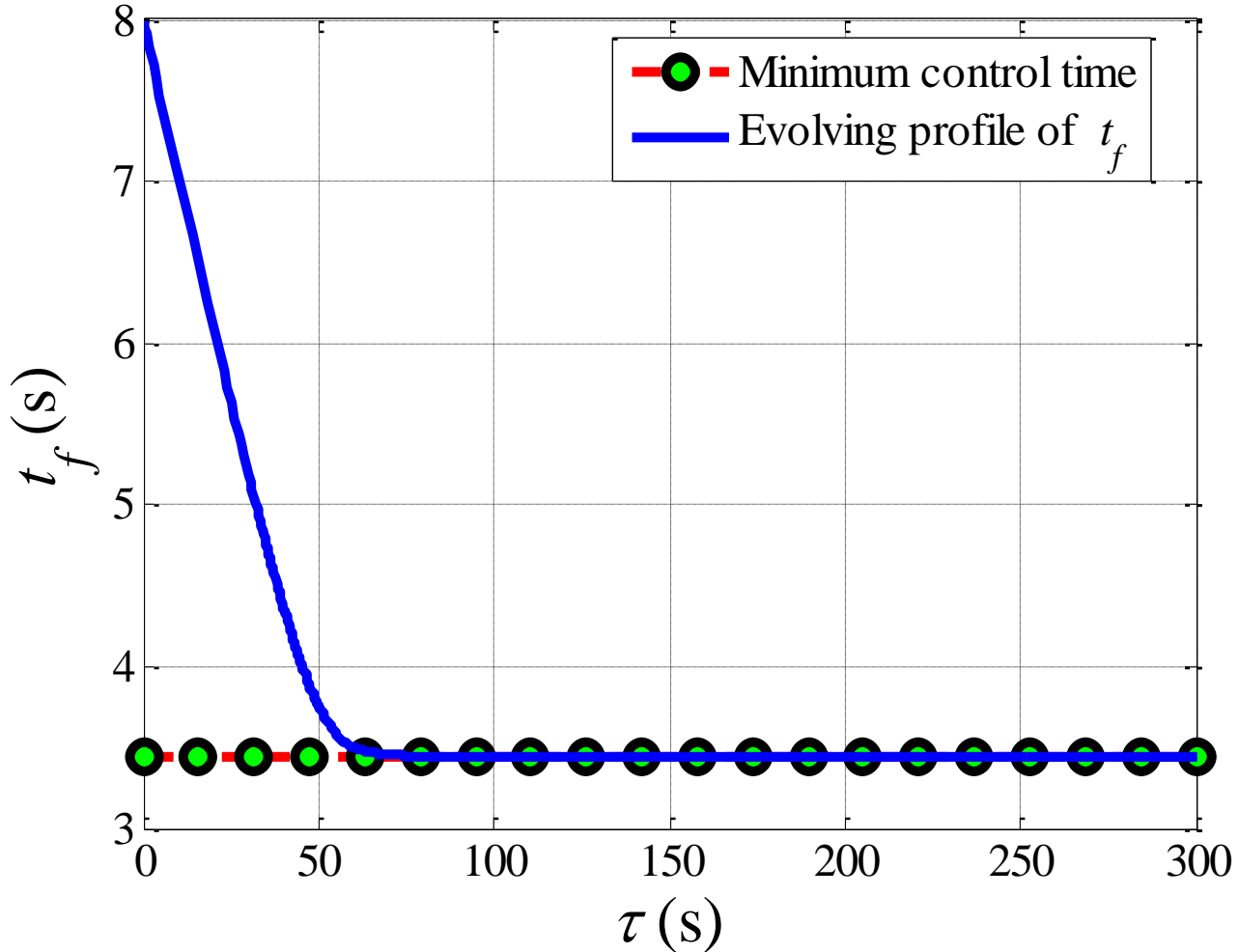

Fig. 5 The evolution profile of $t_f$ to the analytic result.

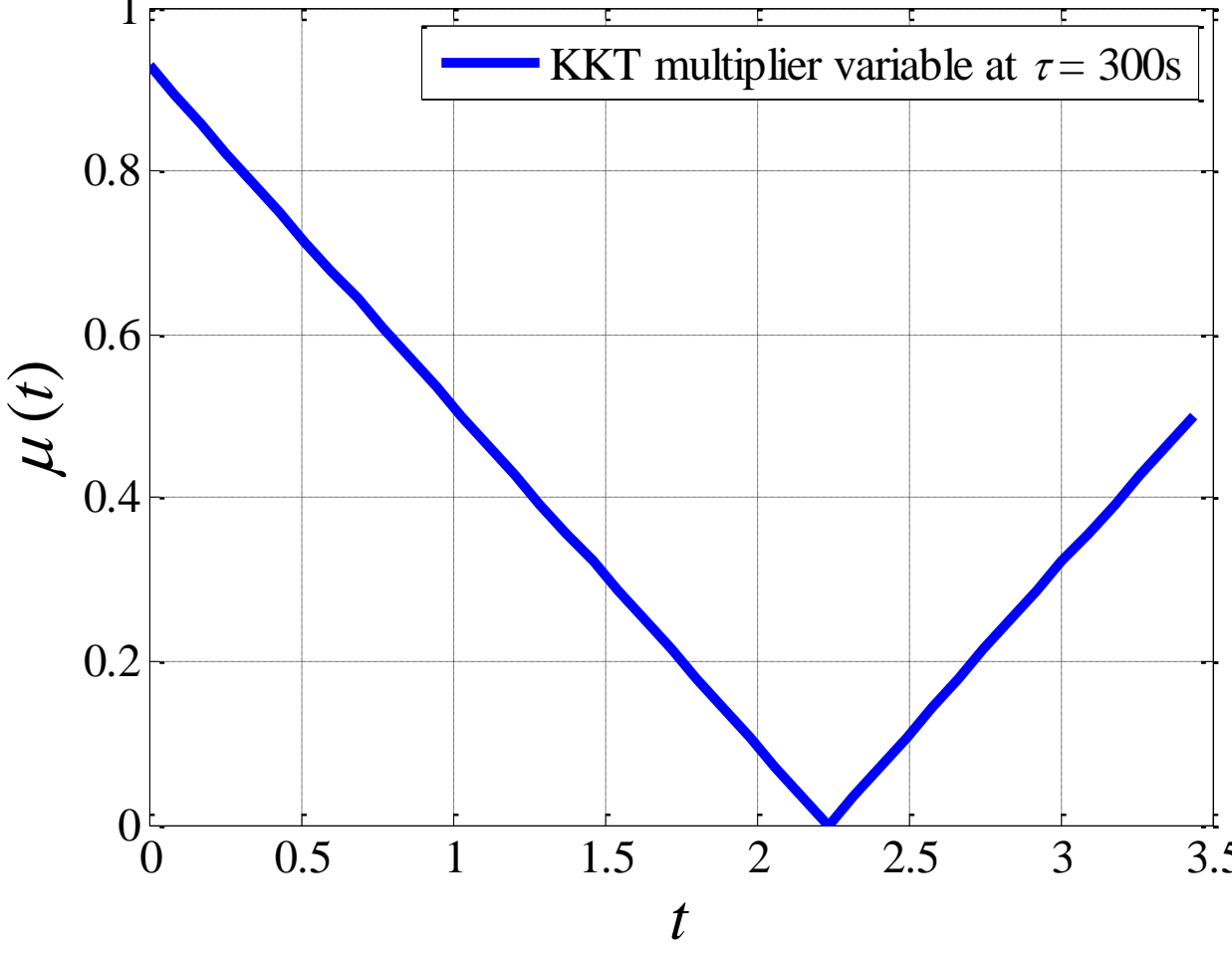

Fig. 6 The KKT multiplier variable profile for the optimal solution with VEM.

Now we consider a nonlinear example with pure-state path constraint, the constrained Brachistochrone problem [28], which describes the motion curve of the fastest descending with position constraint.

**Example 2**: Consider the following dynamic system

$$\dot{\boldsymbol{x}} = \boldsymbol{f}(\boldsymbol{x}, u)$$

where $\boldsymbol{x} = \begin{bmatrix} x \\ y \\ V \end{bmatrix}$, $\boldsymbol{f} = \begin{bmatrix} V\sin(u) \\ -V\cos(u) \\ g\cos(u) \end{bmatrix}$, and $g = 10$ is the gravity constant. Find the solution that minimizes the performance index

$$J = t_f$$

with the boundary conditions

$$\begin{bmatrix} x \\ y \\ V \end{bmatrix}\Bigg|_{t_0=0} = \begin{bmatrix} 0 \\ 0 \\ 0 \end{bmatrix}, \; x\big|_{t_f} = 2$$

In addition, during the descending, the position states are constrained by a slope as

$$C\big(x(t), y(t)\big) = -0.5x(t) - y(t) - 0.35 \le 0$$

In the specific form of the EPDE (103) and the EDE (104), the gain parameters $K$ and $k_{t_f}$ were set to be 0.1 and 0.05, respectively. The barrier parameter $k_C$ in Eq. (120) was set to be 0.2. The definite conditions, i.e., $\begin{bmatrix} \boldsymbol{x}(t,\tau) \\ u(t,\tau) \\ t_f(\tau) \end{bmatrix}\Bigg|_{\tau=0}$, were obtained from a physical motion along a straight line that connects the initial position to the terminal position of $\begin{bmatrix} 2 \\ -1 \end{bmatrix}$, i.e.

$$\begin{aligned} &\tilde{t}_f = 1 \qquad \tilde{u} = \arctan(2) \\ &\tilde{x} = 2t^2 \qquad \tilde{y} = -t^2 \qquad \tilde{V} = 2\sqrt{5}t \end{aligned}$$

We also discretized the time horizon $[t_0, t_f]$ uniformly, with 101 points. Thus, a large IVP with 405 states (including the terminal time) was obtained. We still employed "ode45" in Matlab for the numerical integration. In the integrator setting, the default relative error tolerance and the absolute error tolerance were $1\times10^{-3}$ and $1\times10^{-6}$, respectively. For comparison, we computed the optimal solution with GPOPS-II [31], a Radau PS method based OCP solver.

Fig. 7 gives the states curve in the $xy$ coordinate plane, showing that the numerical results starting from the straight line approach the optimal solution over time, and the optimal descending curve is constrained by the slope. The control solutions are plotted in Fig. 8. The asymptotical approach of the numerical results is demonstrated, and the restriction effect from the slope on the control is clearly shown. In Fig. 9, the terminal time profile against the variation time $\tau$ is plotted. The result of $t_f$ declines rapidly at first and then gradually approaches the minimum decline time, and it only changes slightly after $\tau$ = 50s. At $\tau$ = 300s, we compute that $t_f = 0.8001$s from the VEM, very close to the result of 0.7999s from GPOPS-II. Fig. 10 presents the profiles of the path constraint $C\big(x(t), y(t)\big)$ and the KKT multiplier variable $\mu(t)$ at $\tau$ = 300s, with respect to the $x$ position coordinate. The path constraint is active within the coordinate interval $0.56 \le x \le 1.06$‘, and the corresponding positive KKT multiplier variable is obviously shown. Different from the results in Fig. 6, their values for the active path constraint oscillate due to the numerical error arising from discretization.

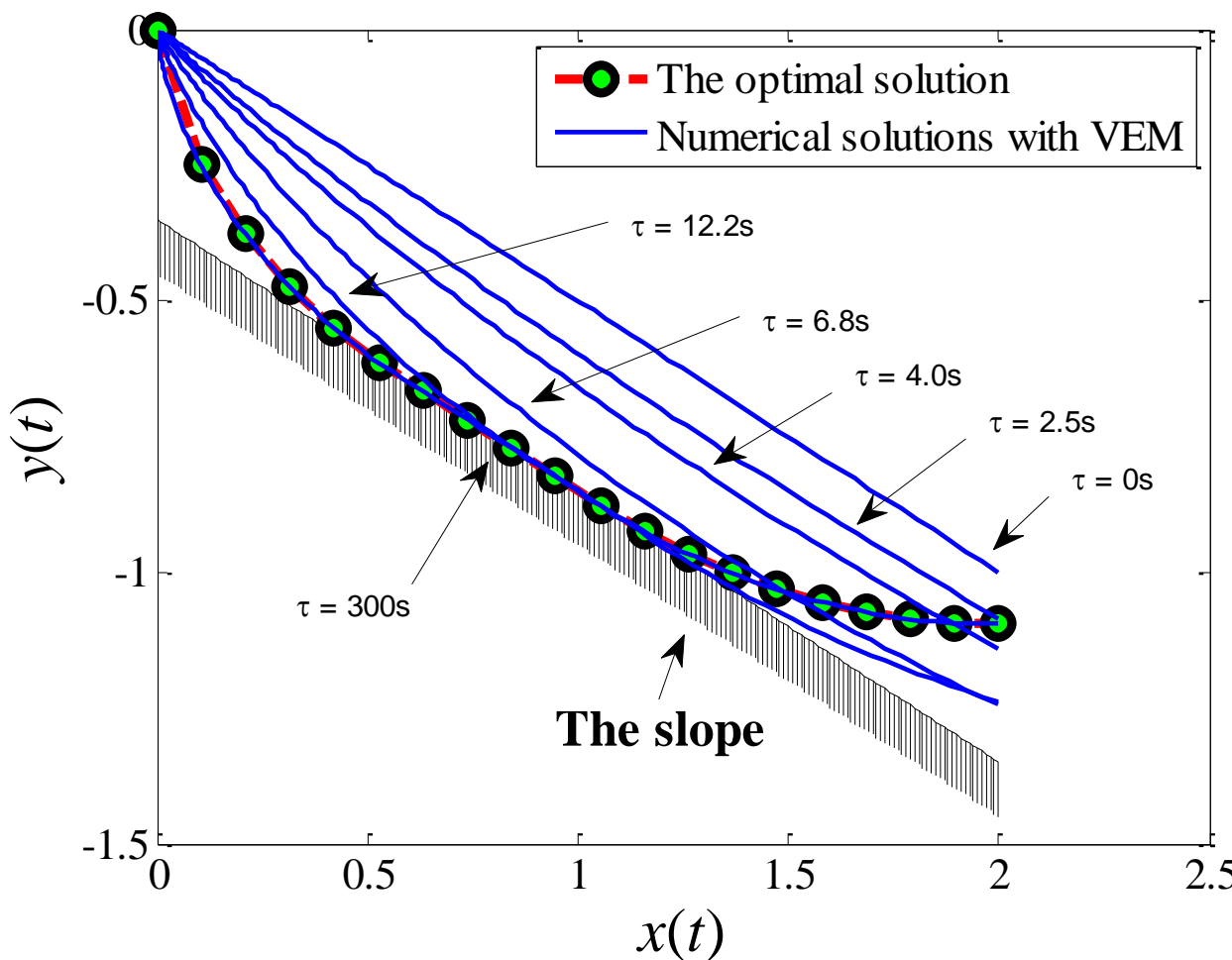

Fig. 7 The evolution of numerical solutions in the $xy$ coordinate plane to the optimal solution.

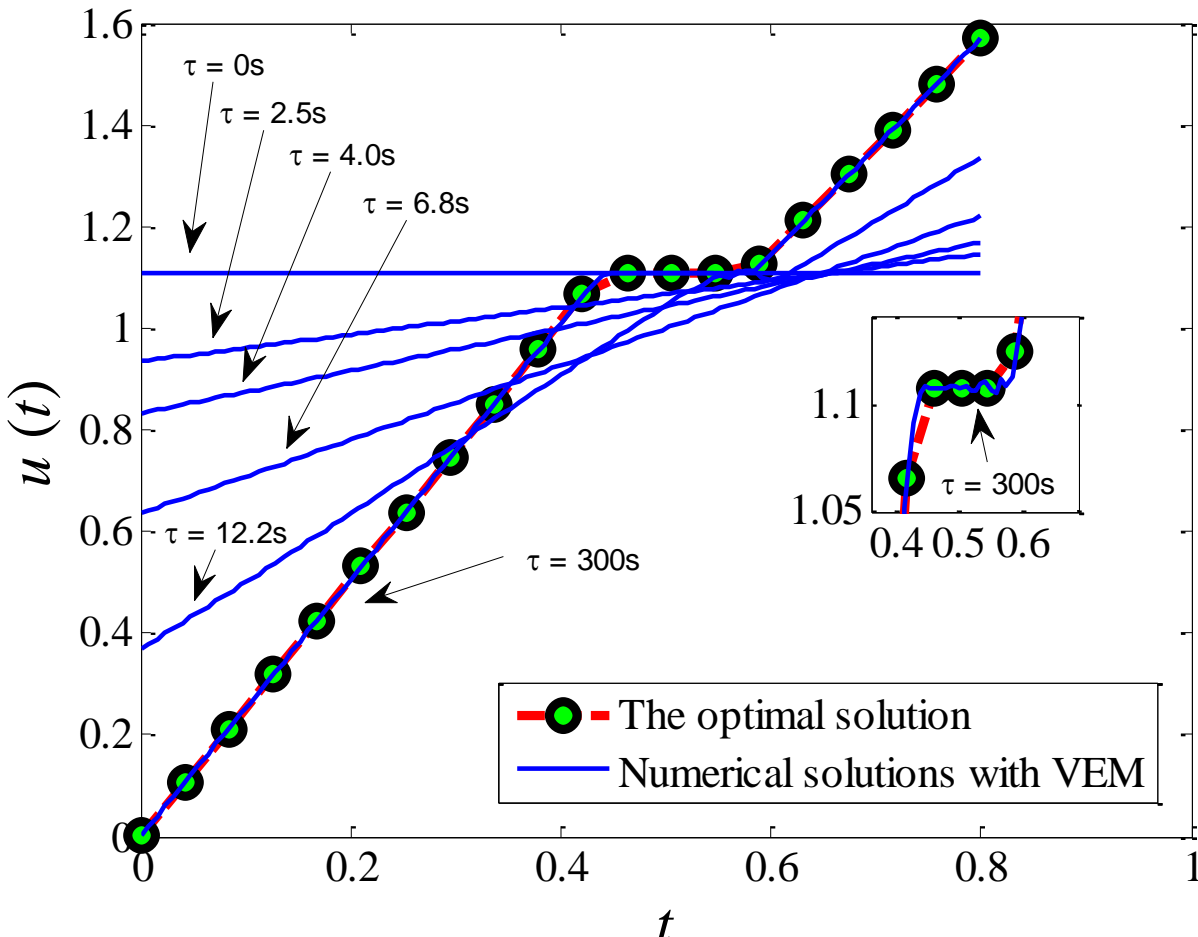

Fig. 8. The evolution of numerical solutions of $u$ to the optimal solution.

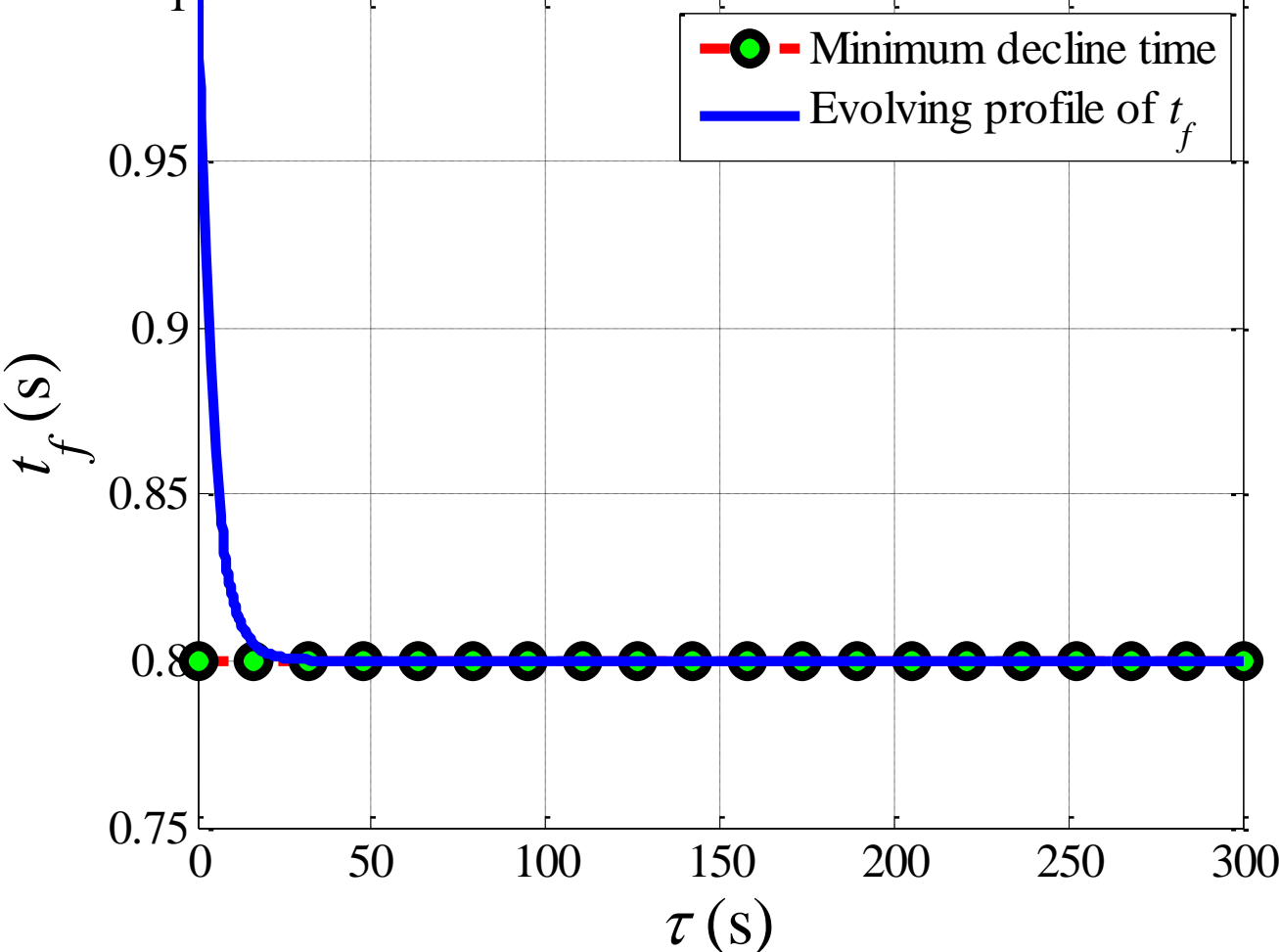

Fig. 9 The evolution profile of $t_f$ to the minimum decline time.

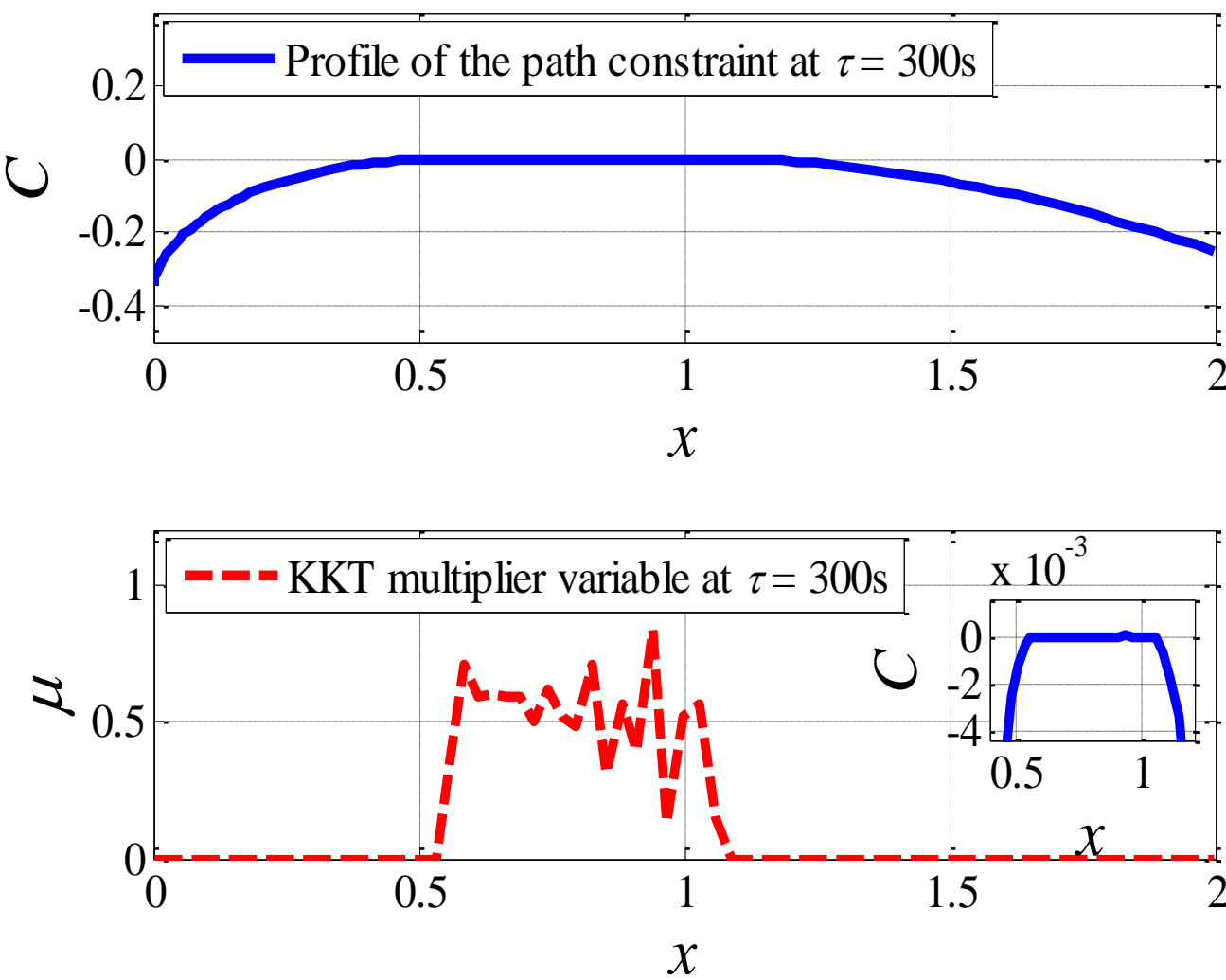

Fig. 10 The profiles of the path constraint and the KKT multiplier with VEM.

## VI. Further Comments

The EPDE derived under the compact VEM is first presented in Ref. [17], and we gave an immature discussion then between it and the augmented EPDE derived in Ref. [15], which was called the first evolution equation originally. Since we have deepened the study along the thread of the compact VEM and achieved systematic results, it makes sense to give a review. To distinguish, the EPDE (103) derived here is named the second evolution equation. For convenience, the first evolution equation is again presented.

$$\frac{\partial \boldsymbol{y}(t,\tau)}{\partial \tau} = -\boldsymbol{K}\left( H_{\boldsymbol{yy}} \begin{bmatrix} \left(H_{\boldsymbol{x}} + \frac{\partial \boldsymbol{\lambda}}{\partial t}\right) \\ \left(\boldsymbol{f} - \frac{\partial \boldsymbol{x}}{\partial t}\right) \\ H_{\boldsymbol{u}} \end{bmatrix} - \frac{\partial}{\partial t}\begin{bmatrix} \left(\frac{\partial \boldsymbol{x}}{\partial t} - \boldsymbol{f}\right) \\ \left(\frac{\partial \boldsymbol{\lambda}}{\partial t} + H_{\boldsymbol{x}}\right) \\ \boldsymbol{0}\end{bmatrix}\right) \tag{121}$$

where $\boldsymbol{y} = \begin{bmatrix} \boldsymbol{x} \\ \boldsymbol{\lambda} \\ \boldsymbol{u} \end{bmatrix}$, $H = L + \boldsymbol{\lambda}^{\mathrm{T}}\boldsymbol{f}$ is the Hamiltonian, and $\boldsymbol{\lambda}$ is the costate variables. $\boldsymbol{K}$ is a $(2n+m)\times(2n+m)$ dimensional positive-definite matrix.

Both the first and the second evolution equations originate from the continuous-time dynamics stability theory, and their solutions are guaranteed to ultimately meet the optimality conditions. The right parts of both equations are only vector functions of time $t$. This makes them suitable to be solved with the semi-discrete method in the field of PDE numerical calculation. Then the numerical solution may be obtained with the common ODE integration methods. However, there exist obvious differences.

The first evolution equation is derived from a constructed unconstrained functional via employing the classic optimality conditions with costates. It may handle typical OCPs with terminal constraint [15]. However, even if it has solved the time-optimal control problem with control constraint [16], it is not applicable to the general constrained OCPs (at least for now). The introduction of the costates also complicates the formula and intensifies the computation burden. In particular, its solution may halt at a saddle point since it cannot differentiate the minimum from the saddle upon their first-order optimality conditions.

The second evolution equation searches the minimum solution from the primary problem, and the equivalent costate-free optimality conditions are established meanwhile, which uncover the analytic relations between the original variables and the augmented quantities, including the costates and the multipliers. It has been shown that the second evolution equation may solve general constrained OCPs and various typical OCPs, and it may be modified to be valid in the infeasible solution domain as Refs. [20] and [21] show. In principle, the second evolution equation requires the integration, and the differentiation, as displayed in the first evolution equation, may be avoided. This is advantageous to reduce the numerical error in seeking optimal solutions.

## VII. CONCLUSION

The Variation Evolving Method (VEM) is developed to solve the general state- and\or control-constrained Optimal Control Problems (OCPs). In deriving the evolution equations, the costate-free optimality conditions are established, and the analytic relations between the original variables and the costates, the KKT multiplier variables, the Lagrange multiplier parameters, introduced in the classic treatment, are uncovered. These results are authenticated between the VEM and the adjoining method, and are helpful to deepen the understanding on the optimal control theory. In our work, the studies of the VEM are carried out upon the assumption that the solution of the OCP exists. Actually this may often be ascertained through the physical analysis. Once the existence of the solution is secured, the VEM theoretically guarantees the convergence to the optimal solution. For the user, this method allows automatically generated initial guess, thus it may be an initial-guess free method for the users. Moreover, the VEM mainly requires common Ordinary Differential Equation (ODE) numerical integration to get the solution. Although we did not highlight the small time consumption, the solutions are usually obtained fairly fast. In its application, since complex numerical computations are avoided, the integration may be implemented with the simple analog circuit. As an outlook, these merits might make the achievement of more reliable and practical on-line optimal control in engineering possible.